%% file: math3.tex
\documentclass[a4paper,11pt,twoside,amsmath,amssymb]{book}

\usepackage[brazil]{babel}
\usepackage[latin1]{inputenc}
\usepackage{indentfirst}
\usepackage{fancyhdr}
\usepackage{graphicx}
\usepackage{epsfig}
\usepackage{subfigure}
\usepackage{float}
\usepackage{array}
\usepackage{multirow}
\usepackage{natbib}
\usepackage{euscript}
\usepackage{stmaryrd}
\usepackage{amsfonts}
\usepackage{amssymb}
\usepackage{amsmath}
\usepackage{hangcaption}
\usepackage{textcomp}
\usepackage{amsthm} 

\newtheorem{definition}{Definição}
\newtheorem{lemma}{Lemma}
\newtheorem{corolary}{Corolário}
\newtheorem*{prova}{Prova}

\newtheorem{prop}{Proposição}

\title{Tópicos em Matemática para as Ciências Mecânicas}
\author{Erick de Moraes Franklin}
\date{\today}

\begin{document}

  
\clearpage
\begin{titlepage}

      \begin{large}
\begin{tabular}[\textwidth]{m{3cm}cm{10cm}}
  \includegraphics[width=3.2cm]{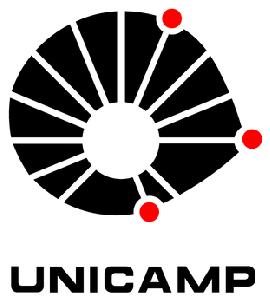} &  &   
      \begin{tabular}{c}
      \emph{Área de concentração: Térmica e Fluidos} \\ 
      \emph{Faculdade de Engenharia Mecânica} \\
      \emph{Universidade Estadual de Campinas - UNICAMP} \\
      \hline
      \end{tabular}
\end{tabular}
      \end{large}
\vspace{4cm}
\begin{center}
      \begin{Huge}
\textbf{Métodos Matemáticos Aplicados à Mecânica dos Fluidos\\}
      \end{Huge}
\vspace{2cm}
\begin{Large}Prof. Dr. Erick de Moraes Franklin\\ \end{Large}
\vspace{2cm}
This manuscript version is made available under the CC-BY-NC-SA 4.0 license https://creativecommons.org/licenses/by-nc-sa/4.0/\\
\vspace{3cm}
\begin{large}
Tópicos abordados: \\
\vspace{1cm}
\begin{tabular}{ll}
1 & Análise Assintótica \\
2 & Funções Generalizadas \\
\end{tabular}

\vspace{2cm}

\end{large}
\end{center}
\end{titlepage}


\clearpage\thispagestyle{empty}\mbox{}\clearpage
\newpage


\pagenumbering{roman}

\clearpage
\chapter*{Preâmbulo}
Este material é baseado nas notas de aula do curso Métodos Matemáticos Aplicados à Mecânica dos Fluidos, lecionado pela primeira vez em 2013 na FEM/UNICAMP. Este curso se propõe a ensinar aos estudantes de Mestrado e Doutorado alguns métodos matemáticos bastante valiosos no tratamento analítico de problemas científicos. Estas ferramentas permitem compreender e analisar o comportamento das equações oriundas de problemas da Física, o que muitas vezes não é possível ser feito através de técnicas numéricas. O objetivo maior destes métodos é auxiliar na real compreensão da física, e não apenas em sua solução. Além disso, veremos que muitos problemas aparentemente impossíveis de serem resolvidos possuem soluções analíticas, principalmente no contexto das distribuições ou funções generalizadas.

Na primeira parte deste curso veremos os fundamentos dos chamados Métodos de Perturbação ou Análise Assintótica. Este método permite, em muitos casos, encontrar soluções tão precisas quanto desejadas para problemas aparentemente sem solução analítica. Esta parte foi inspirada nas notas de aula do Prof. Grégoire Casalis \citep{Casalis} do SUPAERO-Toulouse, entretanto duas referências utilizadas amplamente são os livros do Prof. Milton Van Dyke \citep{VanDyke} e do Prof. Edward John Hinch \citep{Hinch_1}.

A segunda parte do curso trata da Teoria das Distribuições ou Funções Generalizadas conforme elaborada pelo Prof. Laurent Schwartz na década de 1950 \citep{Schwartz_3, Schwartz_1, Schwartz_2}. Esta parte foi inspirada no livro de V.S. Vladimirov \citep{Vladimirov} e no excelente artigo de F. Farassat \citep{Farassat}.

\tableofcontents

\newpage
\thispagestyle{empty}

\newpage
\setcounter{page}{1}

\pagenumbering{arabic}
\include{perturbacao}

\include{distribuicoes}


\addcontentsline{toc}{chapter}{$\mathbf{Bibliografia}$}
\bibliography{references}
\bibliographystyle{plainnat}

\end{document}

%% file: perturbacao.tex
\chapter{Análise Assintótica}
\label{chapter:perturbacao}

\section{Introdução}
\label{section:introduction_perturbacao}

Algumas equações de interesse em Física ou em Matemática Aplicada possuem coeficientes variáveis ou então são não-lineares, de forma que soluções analíticas não podem ser encontradas para muitas destas equações através de métodos usuais em Cálculo.

Entretanto, muitas vezes um pequeno parâmetro (perturbação) está presente na equação. Quando tal parâmetro encontra-se multiplicado por um termo de baixa ordem, muitas vezes a solução é encontrada desprezando-se tal termo. Este é o caso dos problemas de \textit{perturbação regular}. Nos casos em que este pequeno parâmetro encontra-se multiplicado por um termo de alta ordem, então a simplificação da equação pela eliminação deste termo nos conduzirá a uma resposta incompleta. Este é o caso de problemas de \textit{perturbação singular}.

Este capítulo apresenta alguns métodos que permitem encontrar soluções para os casos singulares. Estas soluções são chamadas de \textit{aproximações exatas}, pois nos permitem encontrar aproximações tão boas quanto queiramos e de forma a conhecermos o erro de nossa aproximação (mesmo quando não conhecemos a solução exata!). Nos referimos ao conjunto destes métodos como \textit{Análise Assintótica} ou \textit{Métodos de Perturbação}.

O objetivo de uma análise assintótica é o de determinar o comportamento de uma função bastante complicada comparando esta função, em regiões distintas, com funções conhecidas. Tais funções conhecidas são chamadas de \textit{funções de medição} (ou \textit{gauge functions}, em inglês). Usualmente séries de potência, funções exponenciais e funções logarítmicas são utilizadas como funções de medição. A determinação de quais funções utilizar e em quais regiões é a base da análise assintótica.

Por fim, nota-se aqui que muitas vezes é preferível conhecer o comportamento de uma dada equação perto de fronteiras ou em certas regiões do espaço-tempo ao conhecimento de valores pontuais exatos. Nestes casos, a análise assintótica é uma poderosa ferramenta.

\section{Notação de Landau}

\begin{definition}
\label{definicao_O_1}
Sejam $f$ e $\psi$ duas funções reais ou complexas definidas sobre uma parte $I$ do conjunto dos números reais. Considere que $\psi$ não se anule sobre $I$. Considere ainda que existe uma constante $A\,\in\,\mathbb{R}\,|\,A>0$. Então escrevemos que
	\begin{center}
	$f(x)=O(\psi(x)),\;\;x\,\in\,I$
	\end{center}
\noindent se
	\begin{center}
	$|f(x)/\psi(x)|\,\leq\,A,\;\;\forall x\,\in\,I$
	\end{center}
\end{definition}

\begin{definition}
Sejam as funções $f$ e $\psi$ da Definição \ref{definicao_O_1} e um número real ``$a$" qualquer. Considere um número $\alpha\,\in\,\mathbb{R}\,|\,\alpha>0$. Então escrevemos
	\begin{center}
	$f(x)=O(\psi(x)),\;\;x\,\rightarrow\,a$
	\end{center}
\noindent se
	\begin{center}
	$|f(x)/\psi(x)|\,\leq\,A,\;\;\forall x\,\in\,]a-\alpha ,a+\alpha [$
	\end{center}
\end{definition}

\begin{definition}
Sejam as funções $f$ e $\psi$ da Definição \ref{definicao_O_1}. Considere um número $\alpha\,\in\,\mathbb{R}\,|\,\alpha>0$. Então escrevemos
	\begin{center}
	$f(x)=O(\psi(x)),\;\;x\,\rightarrow\,+\infty$
	\end{center}
\noindent se
	\begin{center}
	$|f(x)/\psi(x)|\,\leq\,A,\;\;\forall x\,\in\,]\alpha ,+\infty [$
	\end{center}
\end{definition}

\begin{definition}
Sejam as funções $f$ e $\psi$ da Definição \ref{definicao_O_1} e um número real ``$a$" qualquer. Então escrevemos
	\begin{center}
	$f(x)=o(\psi(x)),\;\;x\,\rightarrow\,a$
	\end{center}
\noindent se
	\begin{center}
	$\displaystyle\lim_{x\to a}|f(x)/\psi(x)|\,=\,0$
	\end{center}
\end{definition}

\begin{definition}
Sejam as funções $f$ e $\psi$ da Definição \ref{definicao_O_1} e um número real ``$a$" qualquer. Então escrevemos
	\begin{center}
	$f(x)\,\sim\,\psi(x),\;\;x\,\rightarrow\,a$
	\end{center}
\noindent se
	\begin{center}
	$\displaystyle\lim_{x\to a}|f(x)/\psi(x)|\,=\,1$
	\end{center}
	\noindent e dizemos que as funções $f$ e $\psi$ são equivalentes.
\end{definition}

\section{Perturbações singulares e regulares}

O estudo de equações diferenciais lineares nos mostra que há uma relação importante entre estas e relações algébricas. Desta forma, podemos obter informações importantes da análise de perturbações de relações algébricas.  Isto é feito a seguir.

\subsection{Caso regular}
Seja $\epsilon \ll 1$ um pequeno parâmetro (perturbação). Desejamos encontrar a solução da seguinte equação:

\begin{equation}
x^2+\epsilon x-1\,=\,0
\label{eq_regular_1}
\end{equation}

Se fizermos $\epsilon\,=\,0$, encontramos como solução

\begin{equation}
x=-1\;\;e\;\;x=1
\end{equation} 

Procedendo de forma diferente, se calcularmos a solução da Eq. \ref{eq_regular_1} (dado $\epsilon$):

\begin{equation}
x\,=\,\frac{-\epsilon\pm\sqrt{4+\epsilon^2}}{2}
\end{equation}

\noindent obtemos  para $\epsilon\,\rightarrow\,0$ a mesma solução: $x\,=\,\pm 1$

Este é o caso regular: há continuidade das raízes em relação a $\epsilon$. Isto ocorre porque a perturbação não está multiplicada pelo termo de mais alta ordem.

O mesmo ocorre com relação a equações diferenciais.

\subsection{Caso singular}
Seja $\epsilon \ll 1$ um pequeno parâmetro (perturbação). Desejamos encontrar a solução da seguinte equação:

\begin{equation}
\epsilon x^2+x-1\,=\,0
\label{eq_singular_1}
\end{equation}

Neste caso, se fizermos $\epsilon\,=\,0$ encontramos apenas uma raiz: $x=1$. Por outro lado, se calcularmos a raiz da equação \ref{eq_singular_1} (dado $\epsilon$)

\begin{equation}
x\,=\,\frac{-1\pm\sqrt{1+4\epsilon }}{2 \epsilon}
\end{equation}

\noindent obtemos  para $\epsilon\,\rightarrow\,0$:

\begin{equation}
x=-\infty\;\;e\;\;x=1
\end{equation} 

Este é o caso singular: não há continuidade das raízes em relação a $\epsilon$. Isto ocorre porque a perturbação está multiplicada pelo termo de mais alta ordem.

O mesmo ocorre com relação a equações diferenciais.

\section{Método dos desenvolvimentos assintóticos recobertos (MDAR)}
\label{section:mdar}

Este é um método aplicado a equações diferenciais (ordinárias ou parciais) singulares. Este método é bastante útil quando tais perturbações geram equações do tipo \textit{camada limite}.  Este tipo de equação se caracteriza por uma rápida variação em determinada(s) região(ões). Tal região pode ocorrer tanto nas condições de contorno (nas fronteiras) como no interior do intervalo considerado. O exemplo mais conhecido é o da camada limite proveniente das equações de Navier-Stokes. Entretanto, em muitos problemas a camada limite não precisa ter um significado físico, sendo meramente uma região de variação rápida da equação tratada.

\subsection{Exemplo de camada limite em uma fronteira x=0}

Seja a EDO:

\begin{equation}
\left\{ \begin{array}{c} \epsilon y''+(1+x)y'+y=0\\ \, \\ y(0)=1\;\;\;\;y(1)=1 \\ \end{array}\right.
\label{eq_mdar_1}
\end{equation}

Se fizermos $\epsilon =0$ só poderemos atender uma condição limite, e o problema estará mal posto. Percebe-se assim que a perturbação é singular (ela é multiplicada pelo termo de mais alta ordem).

Se resolvermos numericamente esta equação, perceberemos que ela possui uma região de rápida variação, próxima a $x=0$. Entretanto, em geral não conheceremos a região de variação rápida, e deveremos empregar algumas técnicas para tentar prever tal região. Isto será discutido mais para frente.

Temos uma camada limite na região próxima a $x=0$. Uma forma de lidar com ela é empregar o MDAR. Para tanto separamos a equação em duas regiões: uma externa, longe da CL; e uma interna, na CL. Encontraremos uma solução para cada um destes limites assintóticos e depois iremos sobrepô-los (recobrimento).

\subsubsection{Escalas}

Para a região da CL vamos mudar a escala da variável independente. Como nesta região a variação é rápida, devemos ``ampliá-la". Fazemos então $x=\epsilon^pX$, onde $\epsilon \ll 1$ e $p>0$ e definimos $Y(X)=y(x)$. Aplicando esta mudança de variáveis na Eq. \ref{eq_mdar_1}:

\begin{equation}
\epsilon^{1-2p}Y''+\epsilon^{-p}Y'+XY'+Y=0
\label{eq_mdar_2}
\end{equation}

Note que nenhuma aproximação foi feita. Devemos agora encontrar o valor de $p$ para manter o termo de mais alta ordem na equação. O ideal é que este termo permaneça com a mesma ordem de grandeza do termo de ordem (de derivação) imediatamente inferior. No caso da Eq. \ref{eq_mdar_2} temos $p=1$, o que nos leva a:

\begin{equation}
\left\{ \begin{array}{c} Y''+Y'+\epsilon XY'+\epsilon Y=0\\ \, \\ x=\epsilon X \\ \end{array}\right.
\label{eq_mdar_3}
\end{equation}

\subsubsection{Solução externa}

Suporemos que a solução externa é dada por uma série de potências. Dadas as escalas encontradas:

\begin{equation}
y=y_0+\epsilon y_1+\epsilon^2y_2+o(\epsilon^2)
\label{eq_mdar_4}
\end{equation}

\noindent onde $y_n=O(1)$ e com as seguintes condições de contorno:

\begin{equation}
\left\{ \begin{array}{c} y_0(1)=1\\ \, \\ y_n(1)=0,\;\;\forall n>0 \\ \end{array}\right.
\label{eq_mdar_5}
\end{equation}

\noindent e a outra CC será determinada no recobrimento com a região interna.

Inserindo a expansão dada pela Eq. \ref{eq_mdar_4} na Eq. \ref{eq_mdar_1}, obtemos

\begin{equation}
\left\{ \begin{array}{c} (1+x)y_0'+y_0=0\\ \, \\ (1+x)y_1'+y_1=-y_0'' \\ \, \\ (1+x)y_2'+y_2=-y_1'' \\ \end{array}\right.
\label{eq_mdar_6}
\end{equation}

\noindent cujas soluções são:

\begin{equation}
\left\{ \begin{array}{c} y_0=2(1+x)^{-1}\\ \, \\ y_1=(-1/2)(1+x)^{-1}+2(1+x)^{-3} \\ \, \\ y_2=(-1/4)(1+x)^{-1}+(-1/2)(1+x)^{-3}+6(1+x)^{-5} \\ \end{array}\right.
\label{eq_mdar_7}
\end{equation}

A solução externa (truncada na ordem 2) é

\begin{equation}
y=\frac{2}{1+x}+\epsilon \left(\frac{-1}{2(1+x)}+\frac{2}{(1+x)^3}\right) +\epsilon^2 \left( \frac{-1}{4(1+x)}+\frac{-1}{2(1+x)^3}+\frac{6}{(1+x)^5}\right) + o(\epsilon^2)
\label{eq_mdar_8}
\end{equation}

\subsubsection{Solução interna}

Para a solução interna, utiliza-se a escala ``ampliada'' e busca-se uma expansão assintótica. Supondo que esta expansão é dada em série de potências:

\begin{equation}
Y=Y_0+\epsilon Y_1+\epsilon^2Y_2+o(\epsilon^2)
\label{eq_mdar_9}
\end{equation}

\noindent onde $Y_n=O(1)$ e com as seguintes condições de contorno:

\begin{equation}
\left\{ \begin{array}{c} Y_0(0)=1\\ \, \\ Y_n(0)=0,\;\;\forall n>0 \\ \end{array}\right.
\label{eq_mdar_10}
\end{equation}

Inserindo a expansão dada pela Eq. \ref{eq_mdar_9} na Eq. \ref{eq_mdar_3}, obtemos

\begin{equation}
\left\{ \begin{array}{c} Y_0''+Y_0'=0\\ \, \\ Y_1''+Y_1'=-XY_0'-Y_0 \\ \, \\ Y_2''+Y_2'=-XY_1'-Y_1 \\ \end{array}\right.
\label{eq_mdar_11}
\end{equation}

\noindent cujas soluções são:

\begin{equation}
\left\{ \begin{array}{c} Y_0=1+A_0(e^{-X}-1)\\ \, \\ Y_1=-X+A_0(-\frac{1}{2}X^2e^{-X}+X)+A_1(e^{-X}-1) \\ \, \\ Y_2=X^2-2X+A_0(\frac{1}{8}X^4e^{-X}-X^2+2X)+A_1(-\frac{1}{2}X^2e^{-X}+X)+A_2(e^{-X}-1) \\ \end{array}\right.
\label{eq_mdar_12}
\end{equation}

\noindent e onde $A_0$, $A_1$ e $A_2$ são constantes de integração a serem determinadas pelas condições de recobrimento. A solução interna é então:

\begin{equation}
\begin{array}{c} Y= 1+A_0(e^{-X}-1) + \epsilon\left[ -X+A_0(-\frac{1}{2}X^2e^{-X}+X)+A_1(e^{-X}-1)\right] + \\ \epsilon^2 \left[ X^2-2X+A_0(\frac{1}{8}X^4e^{-X}-X^2+2X)+A_1(-\frac{1}{2}X^2e^{-X}+X)+A_2(e^{-X}-1)\right] + \\ + o(\epsilon^2) \end{array}
\label{eq_mdar_13}
\end{equation}

\subsubsection{Recobrimento}

Nas subseções anteriores foram obtidos desenvolvimentos assintóticos para a região externa (Eq. \ref{eq_mdar_8}) e para a região interna (Eq. \ref{eq_mdar_13}). Entretanto, o que se busca é um desenvolvimento assintótico que valha para todo o domínio do problema. Para tanto, é necessário que os desenvolvimentos interno e externo possuam uma região de recobrimento. Neste ponto, é importante observar que os desenvolvimentos interno e externo estão relacionados por $x=\epsilon X$.

Em termos técnicos, o recobrimento é feito ``ordem por ordem'' e é baseado nas seguintes hipóteses:

\begin{itemize}
\item $\exists$ desenvolvimento na região interna para $X\,\rightarrow\,+\infty$ (i.e., $X\gg 1$);
\item $\exists$ desenvolvimento na região externa para $x\,\rightarrow\,0$ (i.e., $x\ll 1$)
\end{itemize}

Os valores das constantes são então determinados de forma que os desenvolvimentos interno e externo coincidam. A região de validade do recobrimento pode ser determinada comparando-se as ordens dos termos desprezados em cada etapa do recobrimento.

\textbf{Ordem 0}

A condição $X\gg 1$ nos permite desprezar os termos em exponenciais de $-X$ na solução interna, o que nos fornece em O(0):

\begin{equation}
Y \sim 1-A_0
\label{eq_mdar_14}
\end{equation}

Igualando a Eq. \ref{eq_mdar_14} com o termos de ordem 0 do desenvolvimento externo (Eq. \ref{eq_mdar_8}) com $x\ll 1$ nos fornece $A_0=-1$.

Procede-se agora à determinação da região de validade do recobrimento na ordem 0. A condição $X\gg 1$ implica que $\epsilon\ll x$. Por outro lado, ao igualar a solução interna à externa, desprezamos os termos $-2x + o(x)$ de $y_0$ (basta expandir em série $y_0$ para perceber isto) face à $2$, o que implica que $x\ll 1$. Na ordem 0, a região de recobrimento é válida para $\epsilon\ll x\ll 1$.

\textbf{Ordem 1}

Procede-se da mesma forma, comparando-se agora os termos $y_0 + \epsilon y_1$ com $Y_0 + \epsilon Y_1$, sendo que o valor de $A_0$ já foi determinado. Substituindo $X$ por $x/\epsilon$ na solução interna, e desprezando os termos em exponenciais de $-X$, obtém-se:

\begin{equation}
Y \sim 2- 2x -\epsilon A_1 + o(\epsilon)
\label{eq_mdar_15}
\end{equation}

Expandindo-se a solução externa em série de Taylor no entorno de $x=0$:

\begin{equation}
y = 2 - 2x + o(x) +\epsilon \left( \frac{3}{2}-\frac{11}{2}x+o(x) \right) + o(\epsilon)
\label{eq_mdar_16}
\end{equation}

Quanto à região de validade do recobrimento, é de se esperar que ela seja mais restrita que a da ordem 0, logo ela deve estar inclusa em $\epsilon\ll x\ll 1$. Ainda, fazer corresponder as Eqs. \ref{eq_mdar_15} e \ref{eq_mdar_16} significa desprezar os termos $x^2$ face aos temos em $\epsilon$, logo $x\ll\sqrt{\epsilon}$. Temos então que em O(1) a região de validade do recobrimento é $\epsilon\ll x\ll \epsilon^{1/2}$. Dentro deste domínio, fica claro que $A_1=-\frac{3}{2}$.

\textbf{Ordem 2}

Compara-se agora os termos $y_0 + \epsilon y_1 + \epsilon^2 y_2$ com $Y_0 + \epsilon Y_1 + \epsilon^2 Y_2$, sendo os valores de $A_0$ e de $A_1$ conhecidos. Substituindo $X$ por $x/\epsilon$ na solução interna, e desprezando os termos em exponenciais de $-X$, obtém-se:

\begin{equation}
Y \sim \left( (1-A_0)(1-x+x^2) \right) + \epsilon \left( -A_1-2x+2A_0x+xA_1 \right) - \epsilon^2A_2 + o(\epsilon^2)
\label{eq_mdar_17}
\end{equation}

Expandindo-se a solução externa em série de Taylor no entorno de $x=0$:

\begin{equation}
\begin{array}{c} y = 2 - 2x + 2x^2 + o(x^2) +\epsilon \left( \frac{3}{2}-\frac{11}{2}x+\frac{23}{2}x^2+o(x^2) \right) + \\ + \epsilon^2\left( \frac{21}{4}-\frac{113}{4}x+\frac{347}{4}x^2+o(x^2) \right) + o(\epsilon^2) \end{array}
\label{eq_mdar_18}
\end{equation}

Para fazer corresponder a Eq. \ref{eq_mdar_18} à Eq. \ref{eq_mdar_17}, é necessário desprezar os termos em $x^3$, $\epsilon x^2$ e $\epsilon^2 x$ do desenvolvimento externo face aos termos $\epsilon^2$ do desenvolvimento interno. Destes, o mais restritivo é o primeiro, o que fornece $x\ll\epsilon^{2/3}$. O domínio de validade dos termos em O(2) do recobrimento é $\epsilon\ll x\ll \epsilon^{2/3}$. Dentro deste domínio, fica claro que $A_2=-\frac{21}{4}$.

Para o recobrimento, podemos nos contentar de:

\begin{equation}
\begin{array}{c} y = 2 - 2x + 2x^2 + o(x^2) +\epsilon \left( \frac{3}{2}-\frac{11}{2}x+o(x) \right) + \\ + \epsilon^2\left( \frac{21}{4}+o(1) \right) + o(\epsilon^2) \end{array}
\label{eq_mdar_18B}
\end{equation}

\begin{figure}
  \begin{center}
    \includegraphics[width=0.6\columnwidth]{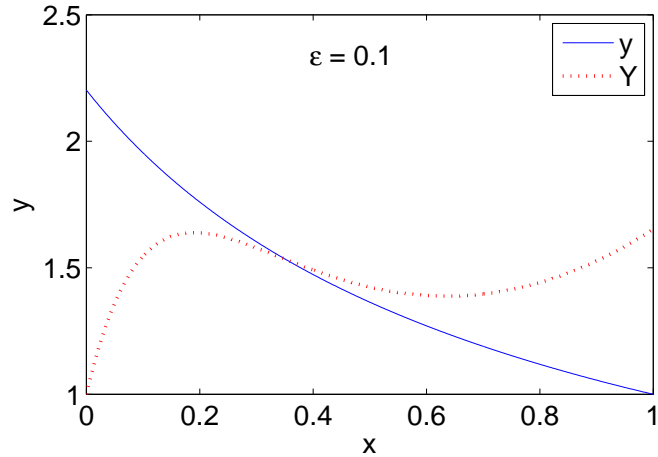}
    \caption{$Y$ e $y$ até $o(\epsilon^2)$ com $\epsilon=0.1$}
    \label{fig:pert_mdar_1}
  \end{center}
\end{figure}

\begin{figure}
  \begin{center}
    \includegraphics[width=0.6\columnwidth]{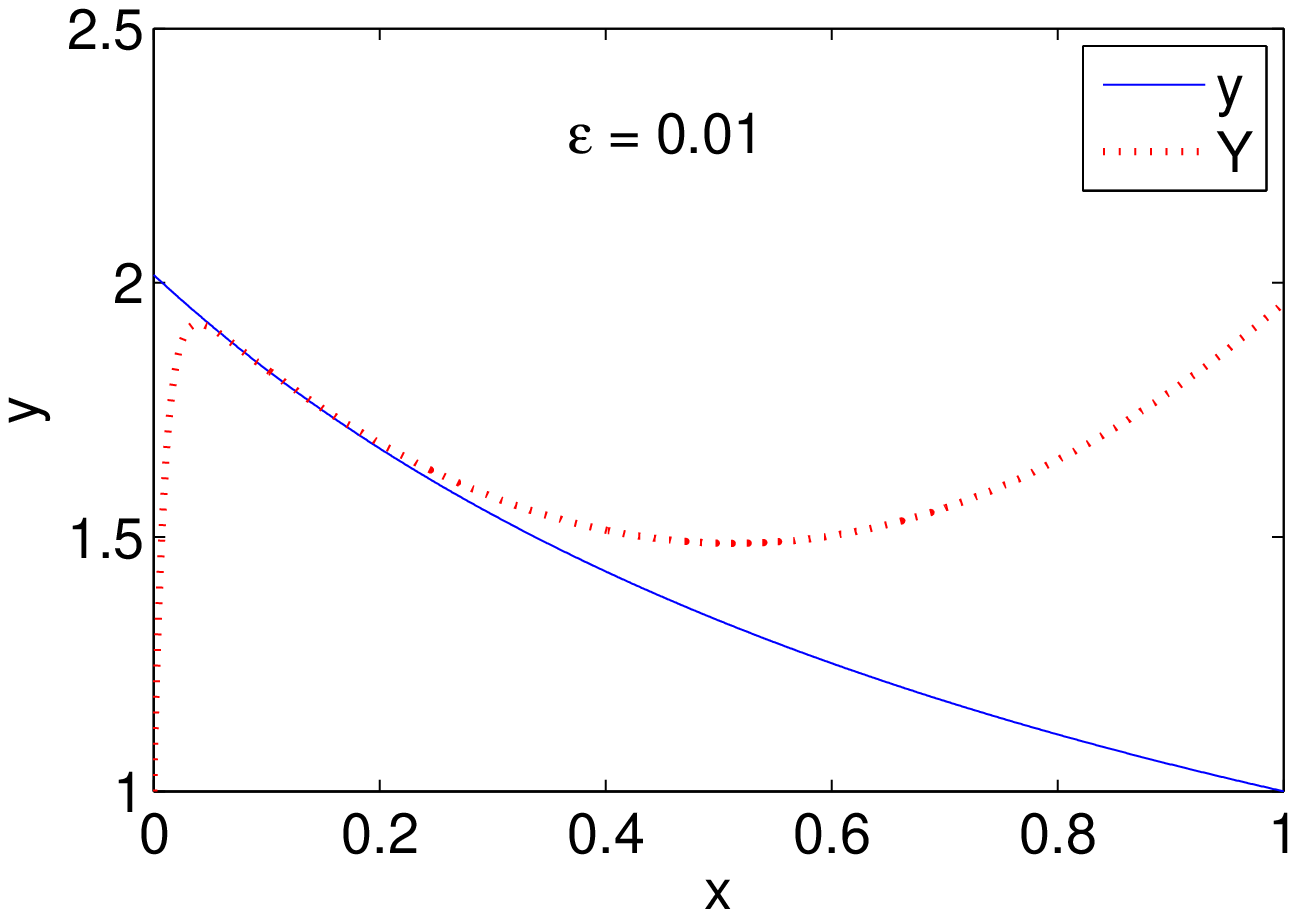}
    \caption{$Y$ e $y$ até $o(\epsilon^2)$ com $\epsilon=0.01$}
    \label{fig:pert_mdar_2}
  \end{center}
\end{figure}

As figuras \ref{fig:pert_mdar_1} e \ref{fig:pert_mdar_2} apresentam os desenvolvimentos interno $Y$ e externo $y$ até $o(\epsilon^2)$ com $\epsilon=0.1$ e $\epsilon=0.01$, respectivamente. Note que, uma vez feito o recobrimento, o desenvolvimento interno ``vale'' da parede até a região de recobrimento, e o desenvolvimento externo ``vale'' da região externa até a região de recobrimento. Note ainda que o recobrimento parece funcionar melhor a medida que o valor de $\epsilon$ diminui (o que está de acordo com a aproximação que buscamos).

\subsubsection{Solução composta}

A solução composta é uma construção, baseada nos desenvolvimentos interno e externo e no conhecimento da região de recobrimento, que seja válida em todo o domínio do problema. Para tanto, basta observar que as constantes foram determinadas baseadas na região de recobrimento, e que temos aproximações válidas tanto para a região ``de recobrimento'' e interna como para a região ``de recobrimento'' e externa. A solução interna se comporta como no recobrimento na região externa e a região externa se comporta como no recobrimento na região interna. Ambas as aproximações possuem o mesmo comportamento na região de recobrimento. Desta forma, a solução composta é dada por:

\begin{equation}
y_c\,=\,y\,+Y\,-\,recobrimento
\label{eq_mdar_19}
\end{equation}

Uma forma alternativa de escrever a solução (e talvez de mais simples compreensão) é a proposta por Van Dyke (\cite{VanDyke})

\begin{equation}
f\,\sim\,\left\{ \begin{array}{c} f_i^m + f_o^n - \left[f_o^n\right]_i^m\\ \, \\ f_i^m + f_o^n - \left[f_i^m\right]_o^n \end{array}\right.
\label{eq_mdar_20}
\end{equation}

\noindent onde $f_i^m$ é o desenvolvimento interno truncado em $m$ termos, $f_o^n$ é o desenvolvimento externo truncado em $n$ termos e $\left[f_o^n\right]_i^m$ são $m$ termos dos $n$ termos do desenvolvimento externo escritos em variáveis internas (e vice-versa).

Para o exemplo em questão, a solução composta pode ser escrita como:

\begin{equation}
y_c\,=\,\left( y_0+Y_0-2\right) + \epsilon\left( y_1+Y_1-\frac{3}{2}+2X\right) + \epsilon^2\left( y_2+Y_2-\frac{21}{4}+\frac{11}{2}X-2X^2\right) + o(\epsilon^2)
\label{eq_mdar_21}
\end{equation}

\begin{figure}
  \begin{center}
    \includegraphics[width=0.6\columnwidth]{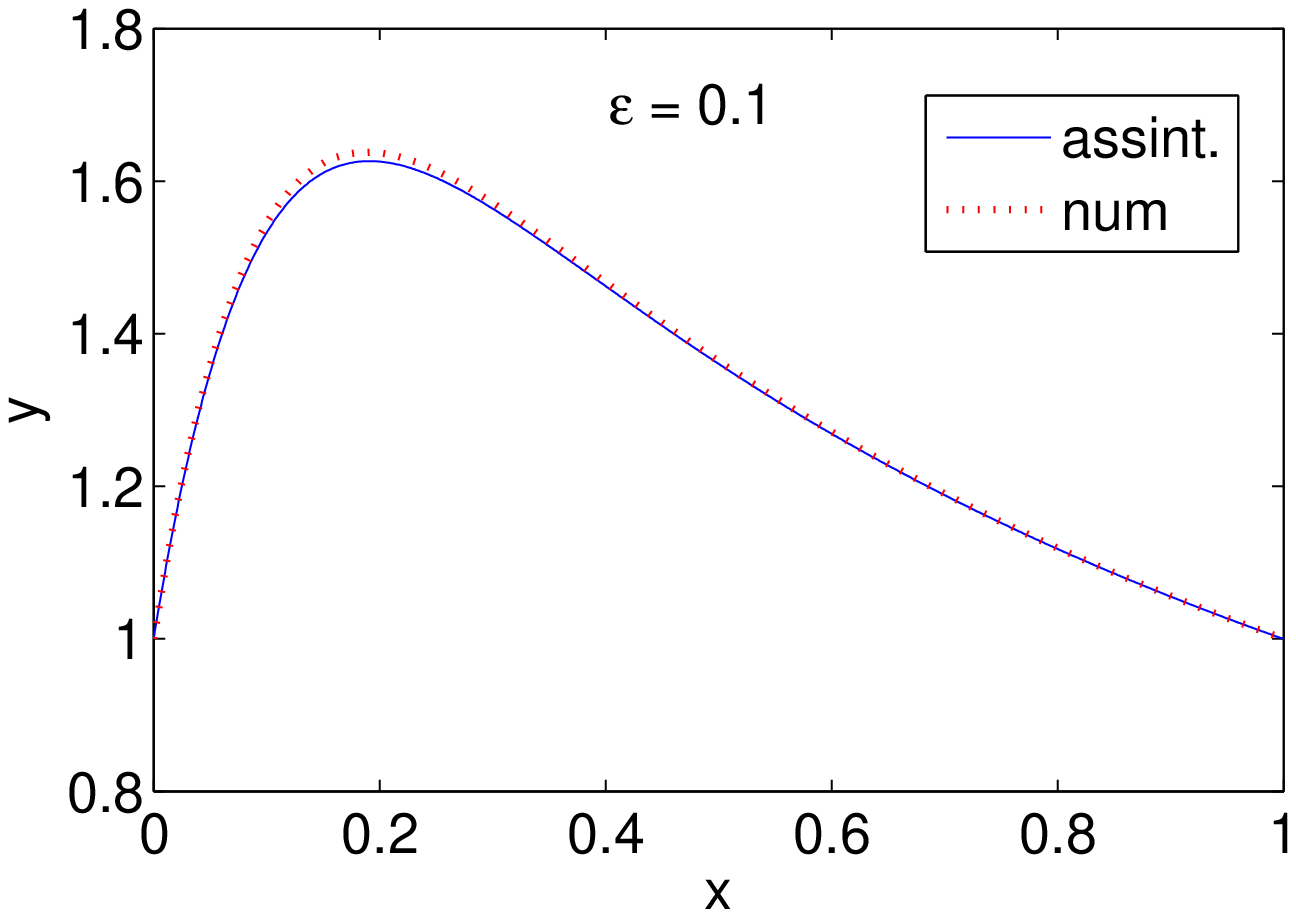}
    \caption{$y_c$ até $o(\epsilon^2)$ e com $\epsilon=0.1$}
    \label{fig:pert_mdar_3}
  \end{center}
\end{figure}

\begin{figure}
  \begin{center}
    \includegraphics[width=0.6\columnwidth]{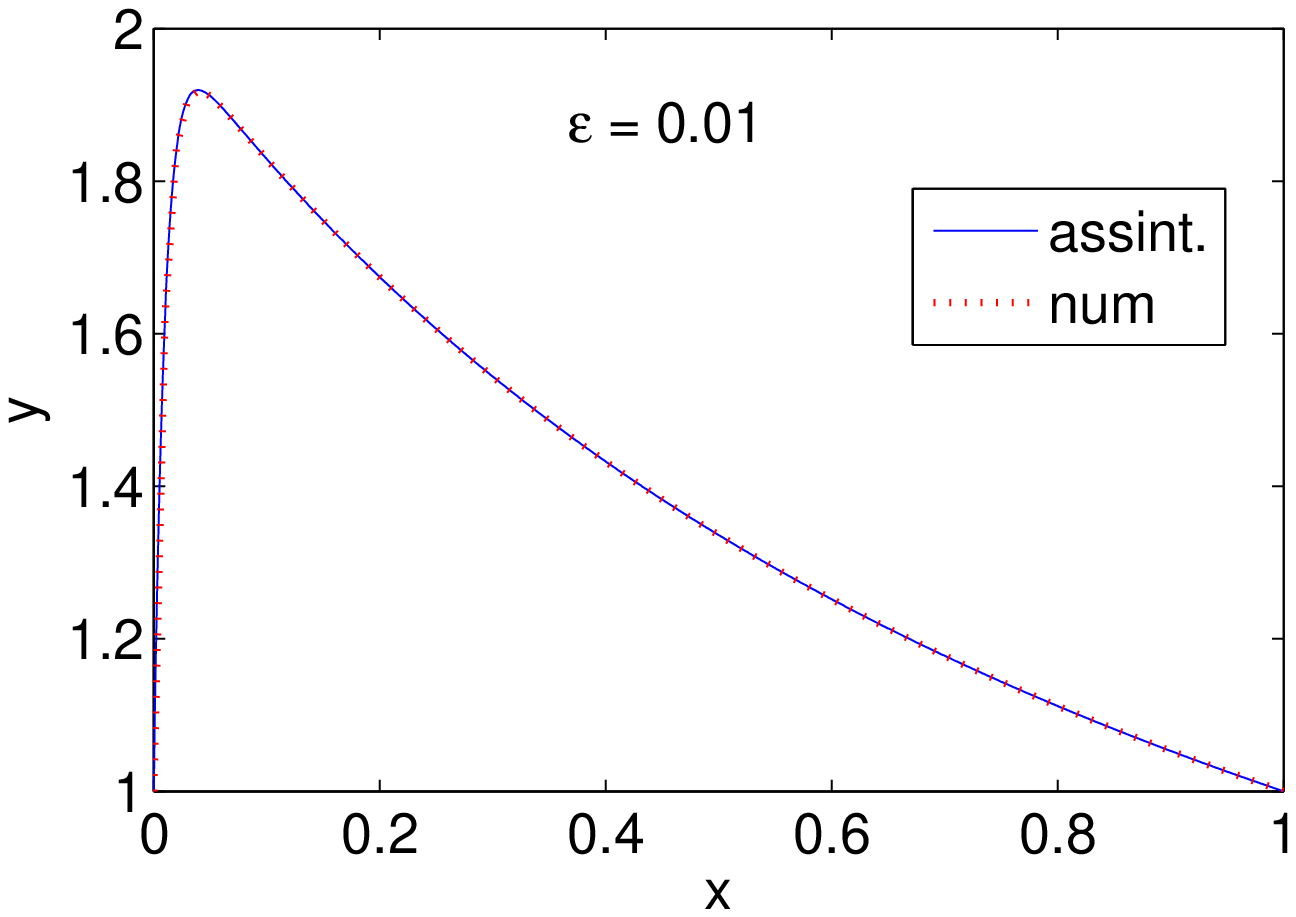}
    \caption{$y_c$ até $o(\epsilon^2)$ e com $\epsilon=0.01$}
    \label{fig:pert_mdar_4}
  \end{center}
\end{figure}

As figuras \ref{fig:pert_mdar_3} e \ref{fig:pert_mdar_4} apresentam a solução composta $y_c$ até $o(\epsilon^2)$ e a solução numérica com $\epsilon=0.1$ e $\epsilon=0.01$, respectivamente. Note que a solução composta se aproxima da solução numérica a medida que o valor de $\epsilon$ diminui.

\subsection{Exemplo: camada limite em x=1}

Considere agora a equação:

\begin{equation}
\left\{ \begin{array}{c} \epsilon y''-(1+x)y'+y=0\\ \, \\ y(0)=1\;\;\;\;y(1)=1 \\ \end{array}\right.
\label{eq_mdar_22}
\end{equation}

Esta equação é quase igual à Eq. \ref{eq_mdar_1}, a única diferença sendo o sinal do termo com derivada de primeira ordem. Aqui, a solução do desenvolvimento externo em ordem zero é:

\begin{equation}
y_0=a\left( 1+x\right)
\label{eq_mdar_23}
\end{equation}

\noindent onde $a$ é uma constante de integração. Se formos buscar uma solução interna em $x=0$, veremos que chegaremos a uma contradição (pois a camada limite está em $x=1$). Assim, se fizermos $x=\epsilon^pX$, onde $\epsilon \ll 1$ e $p>0$, para $Y(X)=y(x)$, e inserirmos isto na Eq. \ref{eq_mdar_22}:

\begin{equation}
\epsilon^{1-2p}Y''-\epsilon^{-p}Y'-XY'+Y=0
\label{eq_mdar_24}
\end{equation}

\noindent onde a única solução não-trivial é obtida quando $p=1$, o que nos leva a:

\begin{equation}
 Y_0''-Y_0'=0
\label{eq_mdar_25}
\end{equation}

\noindent e cuja solução é:

\begin{equation}
 Y_0=b+ce^X
\label{eq_mdar_26}
\end{equation}

Como estamos supondo (erradamente) uma camada limite em $x=0$, a condição em $x=1$ nos fornece $a=1/2$ e as constantes $b$ e $c$ são determinadas das condições em $X=0$ e de recobrimento. Na região de recobrimento, $c=0$ (para que a solução esteja limitada) e $b=1/2$ (para que $y_0(x\rightarrow 0)=Y_0(X\rightarrow\infty)$. Mas, a condição em $X=0$ fornece $b=1$, o que é impossível: logo a hipótese de camada limite em $x=0$ está errada!

Supondo agora que a camada limite se encontra em $x=1$, a condição em $x=0$ (aplicada agora ao desenvolvimento externo) fornece $a=1$. Para a solução interna (na vizinhança de $x=1$), faz-se a seguinte mudança de variáveis: $1-x=\epsilon^pX$, para $Y(X)=y(x)$. Inserindo estas novas variáveis na Eq. \ref{eq_mdar_22}:

\begin{equation}
\epsilon^{1-2p}Y''+2\epsilon^{-p}Y'-XY'+Y=0
\label{eq_mdar_27}
\end{equation}

\noindent onde a única solução não-trivial é obtida quando $p=1$, o que nos leva a:

\begin{equation}
 Y_0''+2Y_0'=0
\label{eq_mdar_28}
\end{equation}

\noindent e cuja solução é:

\begin{equation}
 Y_0=b+ce^{-2X}
\label{eq_mdar_29}
\end{equation}

A condição interna em $X=0$ (ou seja, $x=1$) nos fornece $c=1-b$. A condição de recobrimento $y_0(x\rightarrow 1)=Y_0(X\rightarrow\infty)$ fornece $b=2$. O desenvolvimento interno em O(0) é:

\begin{equation}
 Y_0=2-e^{-2X}
\label{eq_mdar_30}
\end{equation}

E a solução composta é:

\begin{equation}
 y_c= 1 + x -e^{-2(1-x)/\epsilon }
\label{eq_mdar_31}
\end{equation}

\begin{figure}
  \begin{center}
    \includegraphics[width=0.6\columnwidth]{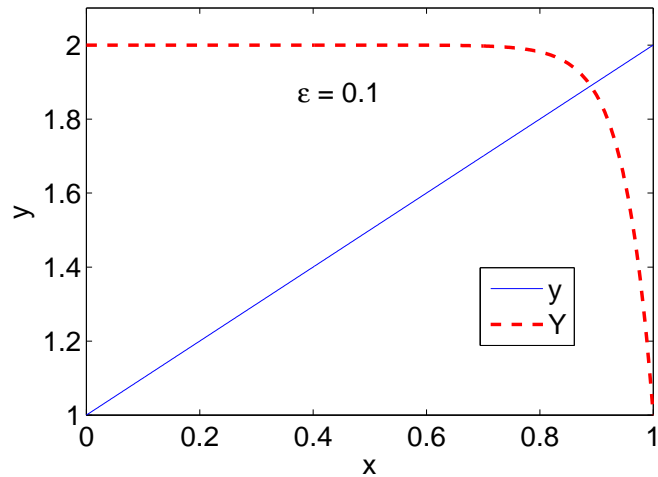}
    \caption{$Y$ e $y$ até $o(1)$ com $\epsilon=0.1$}
    \label{fig:pert_mdar_5}
  \end{center}
\end{figure}

\begin{figure}
  \begin{center}
    \includegraphics[width=0.6\columnwidth]{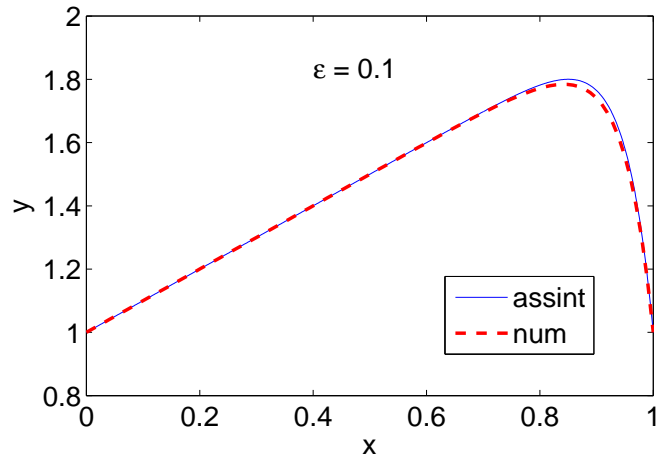}
    \caption{$y_c$ até $o(1)$ e com $\epsilon=0.1$}
    \label{fig:pert_mdar_6}
  \end{center}
\end{figure}

\subsection{Exemplo prático: camada limite hidrodinâmica turbulenta}

The fluid flow close to a wall, in both open and internal flows, has distinct regions. This comes from the fact that its behavior is not the same near the surface, where it is slowed down by viscous effects, and far from the surface, were it is mainly inertial. In between these two regions, there is a matching region. We will focus our analysis in terms of internal flows in channels. However, the same development can be made for external flows.

Far from the wall, the characteristic velocity is the velocity in the center, $U_{c}$, and the characteristic length is the channel height $h_{canal}$. The mean velocity $u$ in this region can be built as a second order correction of the velocity in the center:

\begin{equation}
\frac{u}{U_{c}}\,\sim\,1\,+\Delta F_{1}
\end{equation}

\noindent where $\Delta$ is the first term of a gauge function (then, of order $\epsilon$) and $F_ {1}$ is a function of $Y$ (of order $1$). $Y$ is the coordinate $y$ in terms of external scales:

\begin{equation}
Y\,=\,\frac{y}{h_{canal}}
\end{equation}

In the region near the wall, the flow is slowed down by viscosity. The scales are then small in this region and the viscous effects cannot be neglected. The velocity scale is a small scale $u_{*}$ and the length scale is the viscous length, $\nu/u_{*}$. In this case, the mean velocity $u$ in this region can be considered as proportional to $u_{*}$:

\begin{equation}
\frac{u}{u_{*}}\,\sim\,f_{0}
\end{equation}

\noindent where $f_{0}$ is a function of $y^{+}$, the coordinate $y$ in terms of the internal scales:

\begin{equation}
y^{+}\,=\,\frac{y u_{*}}{\nu}
\end{equation}

As we said, it must exist a matching region between those two regions. If we consider the gauge function as $\Delta\,=\,\frac{u_{*}}{U_{c}}$ and proceed to the matching of the velocities and of their first derivatives, we find the velocity in the matching region. So, doing $Y\,\rightarrow\,0$ and $y^{+}\,\rightarrow\,\infty$, in external scales:

\begin{equation}
\frac{u-U_{c}}{u_{*}}\,=\frac{1}{\kappa}\log{Y}+C_{0}
\label{eq:log_ext}
\end{equation}

\noindent and in internal scales:

\begin{equation}
\frac{u}{u_{*}}\,=\frac{1}{\kappa}\log{y^{+}}+C_{i}
\label{eq:log1}
\end{equation}

\noindent which gives us the well known \textit{log law}. If we write Eq. \ref{eq:log1} with $y_{0}\,=\,\frac{\nu}{u_*}e^{- \kappa C_{i}}$, we find:

\begin{equation}
u\,=\,\frac{u_{*}}{\kappa}\log(\frac{y}{y_{0}})
\label{eq:log2}
\end{equation}

A dimensional analysis with the momentum equation indicates that $u_*=\sqrt{\frac{\tau}{\rho}}$.

\section{Método WKB}
\label{section:WKB}

O Método WKB deve seu nome a 3 físicos que o desenvolveram na década de 1920:  Wentzel, Kramers e Brillouin. Ele é um outro método que se aplica a equações com uma perturbação singular. A particularidade deste método é que ele se aplica apenas a equações lineares (logo razoavelmente limitado), sendo aplicado em problemas lineares envolvendo oscilações de curto comprimento de onda.

Dado o comportamento linear do problema, as soluções podem ser procuradas na formas de somatórios de funções exponenciais. Uma forma adequada é:

\begin{equation}
y(x)\,=\,\exp\left( \frac{1}{\delta} \displaystyle\sum\limits_{n=0}^\infty {\delta^n S_n(x)} \right)
\label{eq_wkb_1}
\end{equation}

\noindent onde $\delta$ é um pequeno parâmetro a ser ajustado e $S_n(x)$ são funções a serem determinadas. As derivadas das soluções da forma da Eq. \ref{eq_wkb_1} podem ser calculadas:

\begin{equation}
y'(x)\,=\,\left( \frac{1}{\delta} \displaystyle\sum\limits_{n=0}^\infty {\delta^n S'_n(x)} \right)\exp\left( \frac{1}{\delta} \displaystyle\sum\limits_{n=0}^\infty {\delta^n S_n(x)} \right)
\label{eq_wkb_2}
\end{equation}

\begin{equation}
y''(x)\,=\,\left[ \left( \frac{1}{\delta} \displaystyle\sum\limits_{n=0}^\infty {\delta^n S''_n(x)} \right) + \left( \frac{1}{\delta} \displaystyle\sum\limits_{n=0}^\infty {\delta^n S'_n(x)} \right)^2 \right]\exp\left( \frac{1}{\delta} \displaystyle\sum\limits_{n=0}^\infty {\delta^n S_n(x)} \right)
\label{eq_wkb_3}
\end{equation}

\noindent e, se nos limitarmos aos 4 primeiros termos do desenvolvimento:

\begin{equation}
\begin{array}{c} y''(x)\,=\, \left[ \frac{1}{\delta^2}S'^2_ 0 + \frac{1}{\delta}\left( S''_0+2S'_0S'_1\right) + \left( S''_1+S'^2_1+2S'_0S'_2\right) + \right. \\ \left. + \delta\left( S''_2+2S'_0S'_3+2S'_1S'_2\right)  + O(\delta^2)\right] y(x) \end{array}
\label{eq_wkb_4}
\end{equation}

\noindent onde $y(x)$ é dado pela Eq. \ref{eq_wkb_1}.

\subsection{Exemplo}

Seja a EDO linear abaixo:

\begin{equation}
\left\{ \begin{array}{c} \epsilon^2 y'' + e^{2x}y=0\\ \, \\ y(0)=1\;\;\;\;y(1)=0 \\ \end{array}\right.
\label{eq_wkb_5}
\end{equation}

Esta é a equação do um oscilador (como veremos) e a técnica do MDAR não pode, a princípio, ser utilizada aqui. Inserindo a expansão dada pela Eq. \ref{eq_wkb_4} na Eq. \ref{eq_wkb_5}, obtemos:

\begin{equation}
\frac{\epsilon^2}{\delta^2}S'^2_ 0 + \frac{\epsilon^2}{\delta}\left( S''_0+2S'_0S'_1\right) + \epsilon^2\left( S''_1+S'^2_1+2S'_0S'_2\right) + o\left(\epsilon^2\right) + e^{2x} = 0
\label{eq_wkb_6}
\end{equation}

\noindent onde uma escolha natural é $\delta = \epsilon$. Na ordem dominante temos 

\begin{equation}
S'^2_ 0 + e^{2x} = 0
\label{eq_wkb_7}
\end{equation}

\noindent cujas soluções são:

\begin{equation}
\left\{ \begin{array}{c} S_{0,(1)}=ie^{x} + a_1 \\ \, \\ S_{0,(2)}=-ie^{x} + a_2 \\ \end{array}\right.
\label{eq_wkb_8}
\end{equation}

\noindent onde $a_1$ e $a_2$ são constantes a serem determinadas. Na ordem seguinte, obtemos a seguinte solução:

\begin{equation}
S_1 = -\frac{1}{2}x+b
\label{eq_wkb_9}
\end{equation}

\noindent onde $b$ é uma constante a ser determinada.

Como a equação é linear, a solução geral pode ser escrita como a soma das contribuições encontradas (Eqs. \ref{eq_wkb_8} e \ref{eq_wkb_9}). É fácil mostrar que ela pode ser escrita na seguinte forma:

\begin{equation}
y = e^{-x/2}\left( \alpha e^{ie^x/\epsilon} + \beta e^{-ie^x/\epsilon} \right)
\label{eq_wkb_10}
\end{equation}

\noindent onde as constantes $\alpha$ e $\beta$ são determinadas das condições de fronteira (Eq. \ref{eq_wkb_5}). A solução é:

\begin{equation}
y = e^{-x/2}\frac{\cos\left( \frac{1-e^x}{\epsilon }\right) - \cos\left( \frac{e^x-2e+1}{\epsilon }\right) }{1- \cos\left( \frac{2-2e}{\epsilon }\right) }
\label{eq_wkb_11}
\end{equation}

\begin{figure}
  \begin{center}
    \includegraphics[width=0.6\columnwidth]{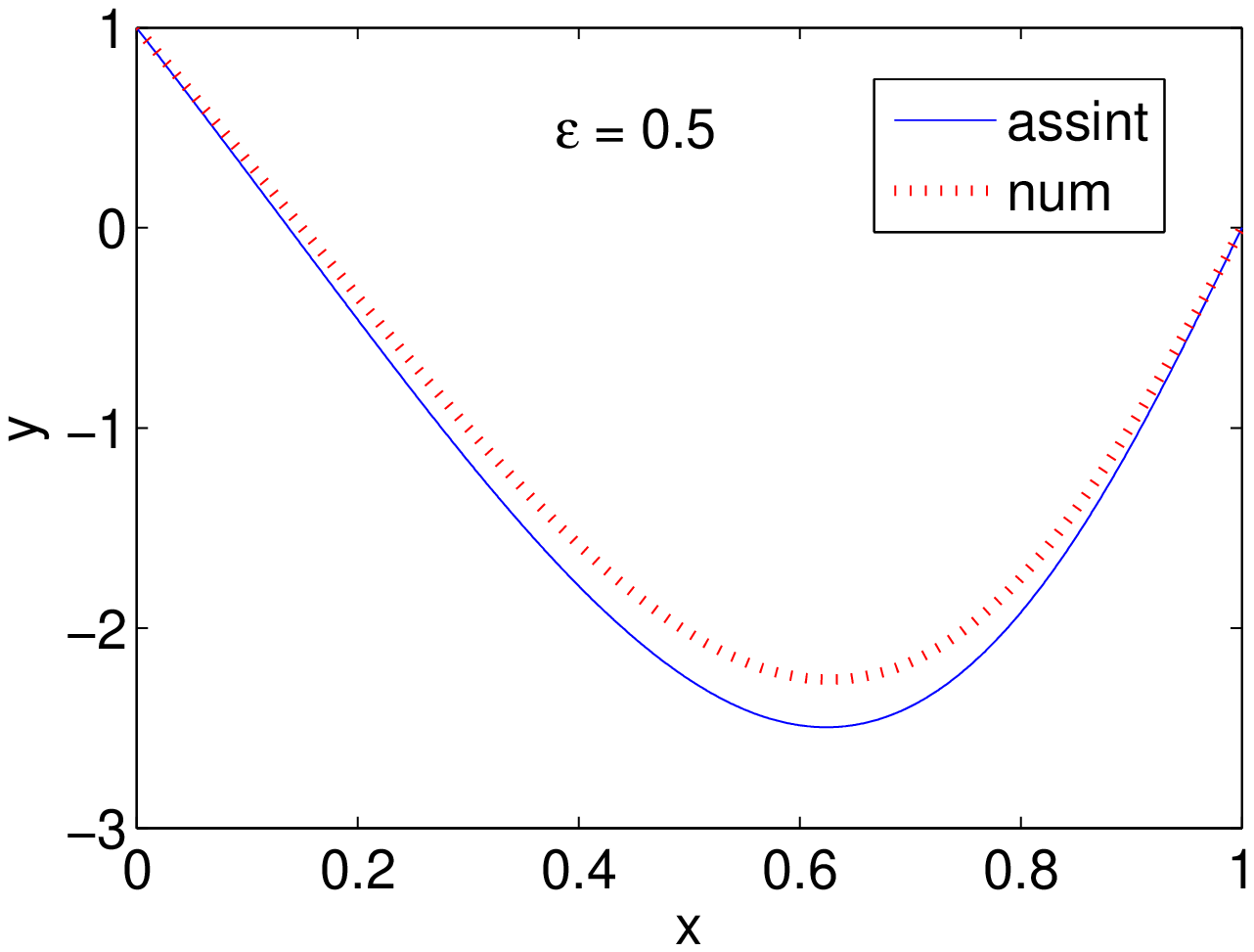}
    \caption{Comparação da solução assintótica até $o(\epsilon^2)$ e com $\epsilon=0.5$ com a solução numérica}
    \label{fig:pert_wkb_1}
  \end{center}
\end{figure}

\begin{figure}
  \begin{center}
    \includegraphics[width=0.6\columnwidth]{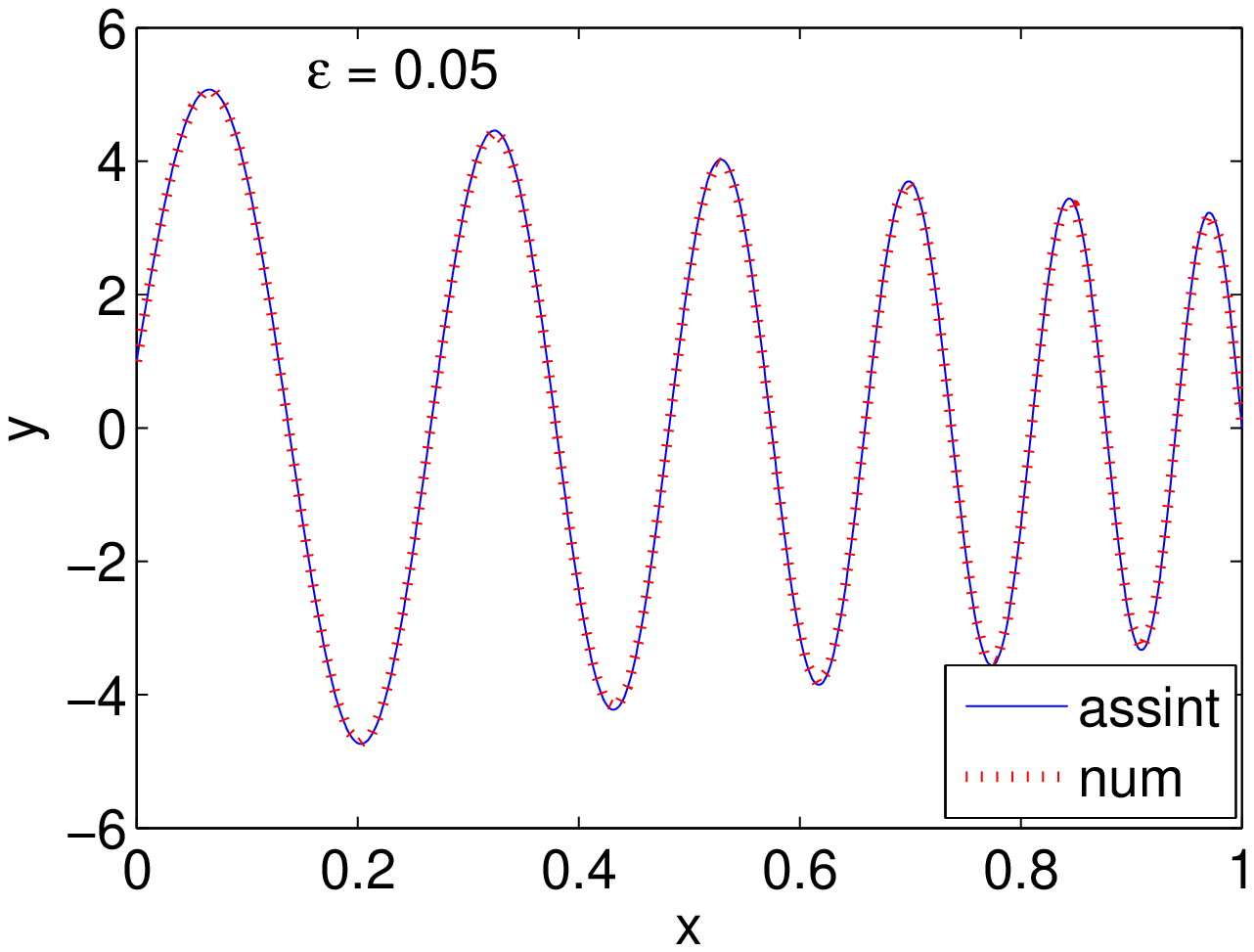}
    \caption{Comparação da solução assintótica até $o(\epsilon^2)$ e com $\epsilon=0.05$ com a solução numérica}
    \label{fig:pert_wkb_3}
  \end{center}
\end{figure}

As figuras \ref{fig:pert_wkb_1} a \ref{fig:pert_wkb_3} comparam a solução assintótica dada pela Eq. \ref{eq_wkb_11} com a solução numérica, para $\epsilon=0.5$, $\epsilon=0.1$ e $\epsilon=0.05$, respectivamente. Podemos perceber que trata-se efetivamente de um oscilados, cuja frequência aumenta com a diminuição de $\epsilon$. Ainda, quanto menor o valor de $\epsilon$, melhor é a aproximação (como esperado).

\begin{figure}
  \begin{center}
    \includegraphics[width=0.6\columnwidth]{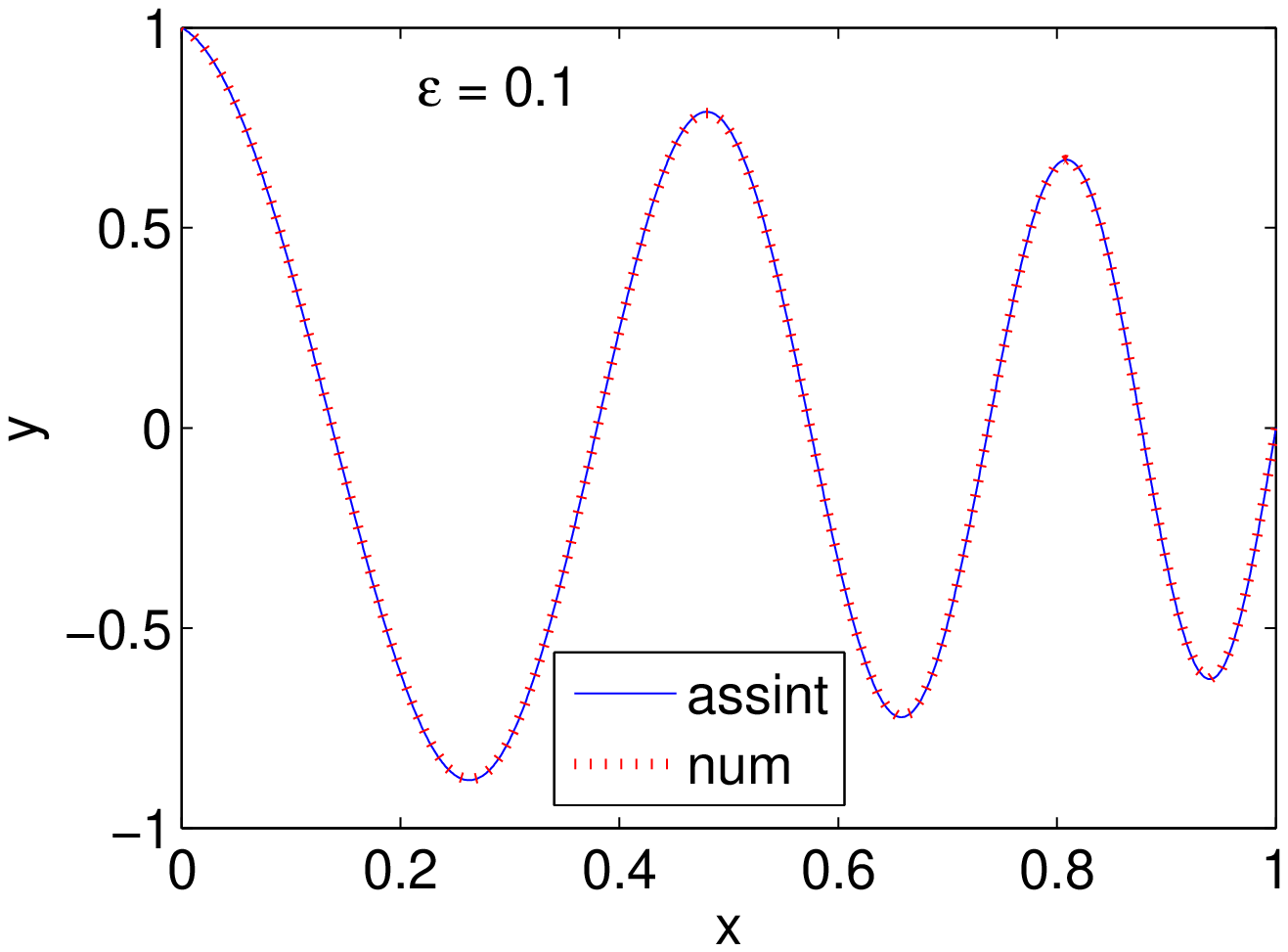}
    \caption{Comparação da solução assintótica até $o(\epsilon^2)$ e com $\epsilon=0.1$ com a solução numérica}
    \label{fig:pert_wkb_2}
  \end{center}
\end{figure}

\section{Método de Múltiplas Escalas (MME)}
\label{section:mult_esc}

Algumas EDOs descrevem o comportamento de osciladores para os quais existem duas escalas de tempo distintas: uma de variação rápida (alta frequência) e outra lenta. Dependendo das ordens de grandeza envolvidas, o termo de variação lenta pode ``demorar'' a aparecer na solução do problema, isto é, em curtos períodos de tempo, percebemos apenas as variações de alta frequência. Entretanto, em períodos longos o efeito do termo lento não pode ser desprezado. Este é o caso típico dos sistemas planetários, e por este motivo o termo lento é usualmente chamado de \textit{termo secular}. Nestes casos, para solucionar corretamente o problema é necessário que as duas escalas de tempo sejam consideradas. Este é o objetivo do Método das Múltiplas Escalas.

\subsection{Exemplo: oscilador de Rayleigh}

Seja a EDO abaixo:

\begin{equation}
\left\{ \begin{array}{c} y'' + y -\epsilon\left( y'-\frac{1}{3}y'^3\right)=0\\ \, \\ y(0)=0\;\;\;\;y'(0)=2a>0 \\ \end{array}\right.
\label{eq_mme_1}
\end{equation}

\noindent onde a variável independente é $t$ (uma vez que trata-se de um oscilador).

Esta equação descreve oscilações rápidas no interior de um envelope lentamente variável (isto pode ser observado resolvendo-se numericamente a equação). Para encontrar uma aproximação, vamos considerar uma escala de tempo rápida $t$ (a variável independente presente na Eq. \ref{eq_mme_1}) e definiremos uma escala de tempo lenta $\tau = \epsilon t$, que também é considerada uma variável independente. Desenvolve-se então a variável dependente como, por exemplo, abaixo:

\begin{equation}
y\,=\, Y_0(t,\tau ) + \epsilon Y_1(t,\tau ) + o(\epsilon )
\label{eq_mme_2}
\end{equation}

\noindent onde $Y_0$ e $Y_1$ são $O(1)$ Se inserirmos o desenvolvimento (Eq. \ref{eq_mme_2}) na Eq. \ref{eq_mme_1}, obtemos em $O(1)$:

\begin{equation}
\frac{\partial^2Y_0}{\partial t^2} + Y_0 = 0
\label{eq_mme_3}
\end{equation}

\noindent cuja solução é:

\begin{equation}
Y_0 = A(\tau )e^{it} + A^*(\tau )e^{-it}
\label{eq_mme_4}
\end{equation}

\noindent onde o índice $*$ representa o complexo conjugado.

Para $O(\epsilon)$, obtemos:

\begin{equation}
\frac{\partial^2Y_1}{\partial t^2} + Y_1 = \frac{\partial Y_0}{\partial t} - \frac{1}{3}\left( \frac{\partial Y_0}{\partial t}\right)^3 - 2\frac{\partial^2Y_0}{\partial t\partial\tau}
\label{eq_mme_5}
\end{equation}

Entretanto, a observação de que $Y_0$ e $Y_1$ são $O(1)$ nos permite obter a partir da Eq. \ref{eq_mme_5} uma condição que nos levará à solução. A substituição da Eq. \ref{eq_mme_4} no lado direito da Eq. \ref{eq_mme_5} fornece uma equação do tipo:

\begin{equation}
\frac{\partial Y_0}{\partial t} - \frac{1}{3}\left( \frac{\partial Y_0}{\partial t}\right)^3 - 2\frac{\partial^2Y_0}{\partial t\partial\tau} = \xi_1(\tau )e^{it} + \xi^*_1(\tau )e^{-it} + \xi_3(\tau )e^{3it} + \xi^*_3(\tau )e^{-3it}
\label{eq_mme_6}
\end{equation}

onde

\begin{equation*}
\left\{ \begin{array}{c} \xi_1(\tau ) = iA(\tau ) - iA^2(\tau )^2 A^*(\tau ) - 2iA'(\tau )\\ \, \\ \xi^*_1(\tau ) = -iA^*(\tau ) + iA(\tau ) A^{*2}(\tau ) + 2iA^{*'}(\tau ) \\ \, \\ \xi_3(\tau ) = \frac{1}{3}iA^3(\tau ) \\ \, \\ \xi^*_3(\tau ) = -\frac{1}{3}iA^{*3}(\tau ) \\ \end{array}\right.
\end{equation*}

Nota-se agora que os termos em $e^{it}$ e em $e^{-it}$ correspondem às frequências naturais do lado esquerdo da Eq. \ref{eq_mme_5}. Como $Y_0$ e $Y_1$ são $O(1)$, estes termos devem se anular (caso contrário, haverá ressonância). Utilizando, por exemplo, $\xi_1(\tau )$, obtemos a Equação de Landau:

\begin{equation}
A(\tau ) - A^2(\tau )A^*(\tau ) - 2A'(\tau )=0
\label{eq_mme_7}
\end{equation}

\noindent e, decompondo-se $A(\tau )$ em módulo $\rho$ e phase $\theta$

\begin{equation}
A(\tau ) = \rho(\tau )e^{i\phi(\tau )}
\label{eq_mme_8}
\end{equation}

\noindent obtemos o sistema de equações

\begin{equation}
\left\{ \begin{array}{c} \theta '=0\\ \, \\ \rho - \rho^3 = 2\rho ' \\ \end{array}\right.
\label{eq_mme_9}
\end{equation}

\noindent cujas soluções são

\begin{equation}
\left\{ \begin{array}{c} \theta = \theta_0 \\ \, \\ \rho = \frac{C}{\sqrt{1+C^2e^\tau }}e^{\tau /2} \\ \end{array}\right.
\label{eq_mme_10}
\end{equation}

\noindent onde $\theta_0$ e $C$ são constantes a serem determinadas. Assim, no espaço real temos:

\begin{equation}
Y_0 = \frac{2C}{\sqrt{1+C^2e^\tau }}e^{\tau /2} \cos (t+\theta_0)
\label{eq_mme_11}
\end{equation}

A condição de contorno $y(0)=0$ nos fornece $\theta_0=\pi /2$ (ou $\theta_0=-\pi /2$, o que dará o mesmo resultado). Já a condição $y'(0)=2a$ nos fornece o valor de $C$. A solução final (em O(1)) é:

\begin{equation}
y(t) = \left[ \frac{2a}{\sqrt{1+a^2(e^\tau -1)}}e^{\tau /2}\right] \sin (t) + O(\epsilon)
\label{eq_mme_12}
\end{equation}

\noindent onde o termo entre colchetes representa o envelope $g$ dentro do qual ocorrem as variações rápidas ($\sin (t)$).

\begin{figure}
  \begin{center}
    \includegraphics[width=0.6\columnwidth]{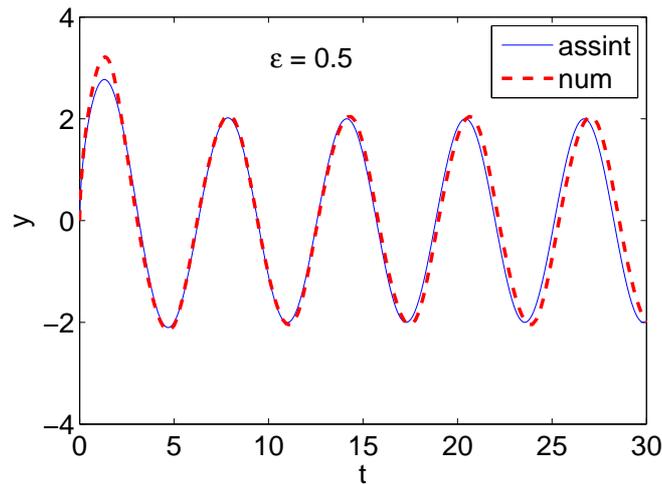}
    \caption{Comparação da solução assintótica até $O(1)$ e com $\epsilon=0.5$ com a solução numérica}
    \label{fig:pert_mme_1}
  \end{center}
\end{figure}

\begin{figure}
  \begin{center}
    \includegraphics[width=0.6\columnwidth]{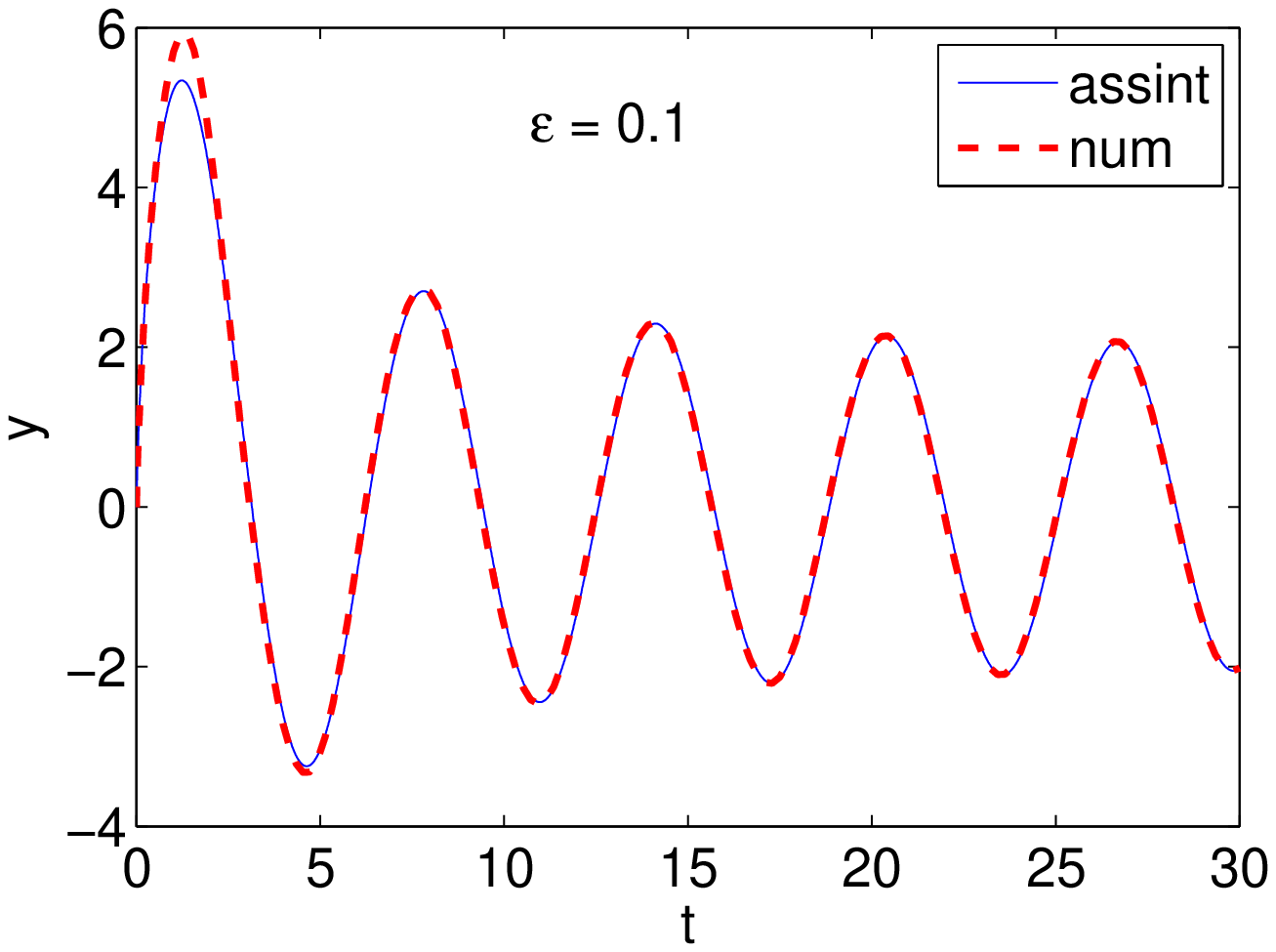}
    \caption{Comparação da solução assintótica até $O(1)$ e com $\epsilon=0.1$ com a solução numérica}
    \label{fig:pert_mme_2}
  \end{center}
\end{figure}

\begin{figure}
  \begin{center}
    \includegraphics[width=0.6\columnwidth]{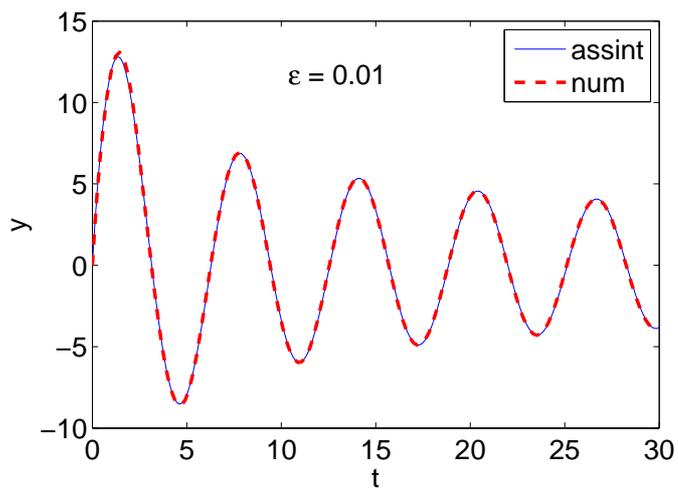}
    \caption{Comparação da solução assintótica até $O(1)$ e com $\epsilon=0.01$ com a solução numérica}
    \label{fig:pert_mme_3}
  \end{center}
\end{figure}

\begin{figure}
   \begin{minipage}[c]{.49\linewidth}
    \begin{center}
      \includegraphics[scale=0.45]{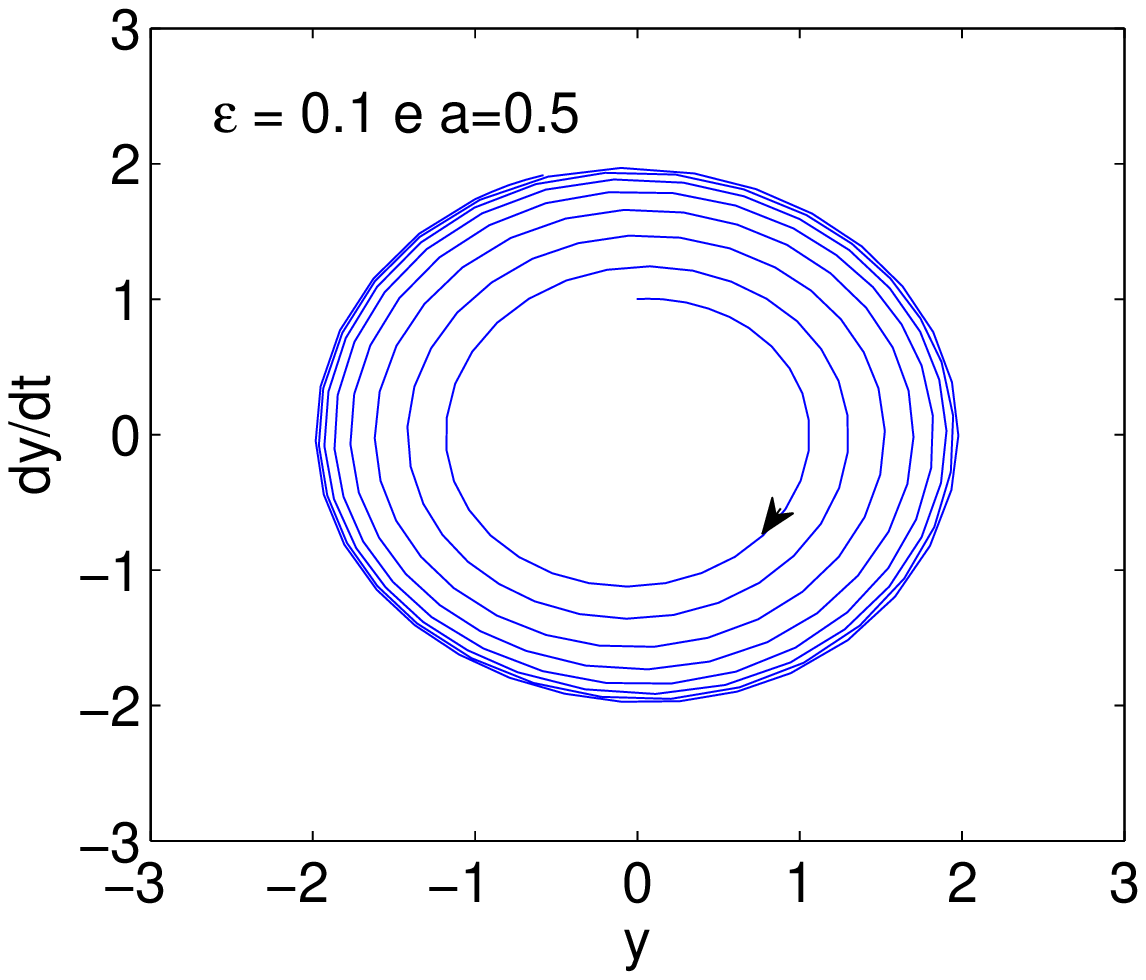}
    \end{center}
   \end{minipage} \hfill
   \begin{minipage}[c]{.49\linewidth}
    \begin{center}
      \includegraphics[scale=0.45]{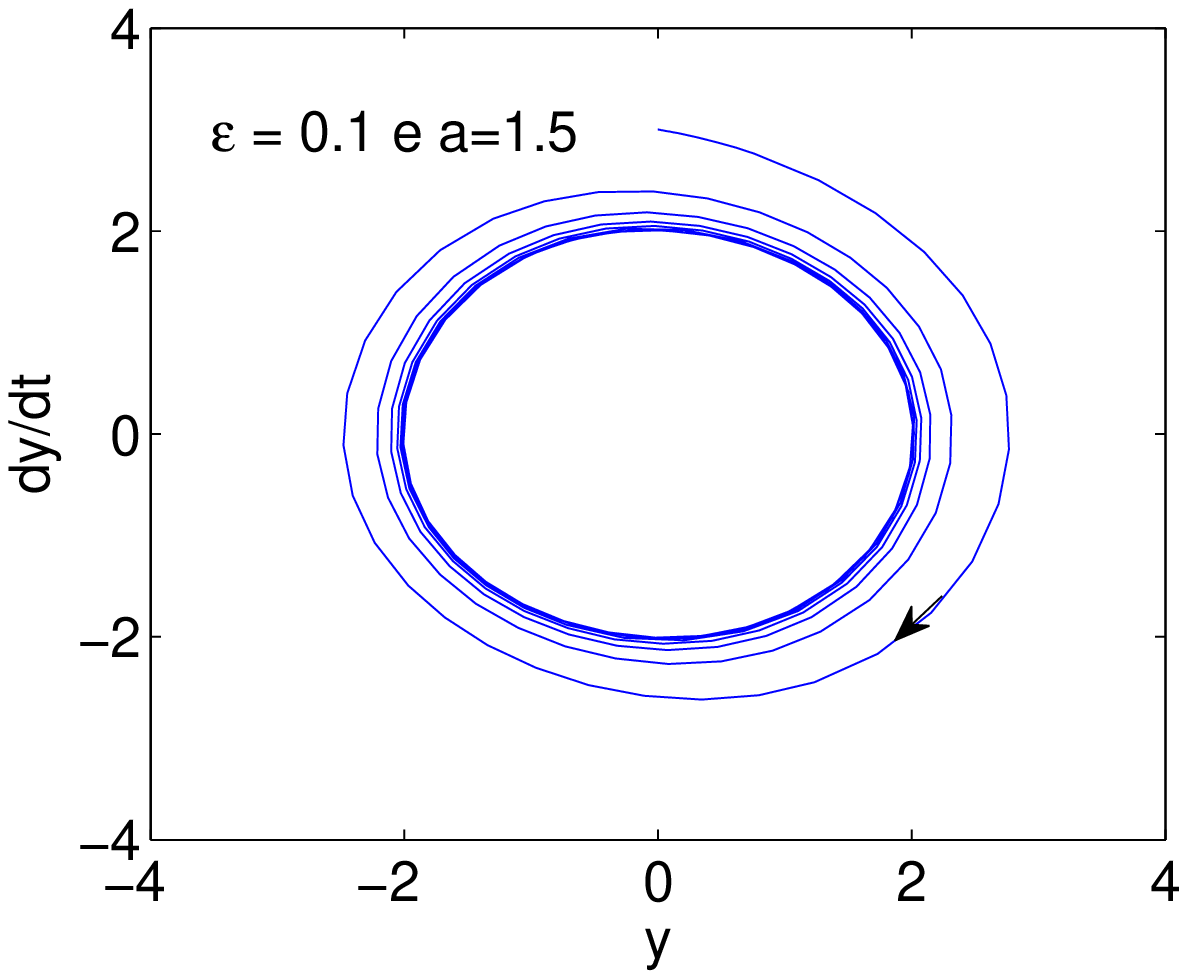}
    \end{center}
   \end{minipage}
\caption{Plano de fases}
\label{fig:pert_mme_4}
\end{figure}

É interessante notar que quando $t\rightarrow\infty$ (ou $\tau\rightarrow\infty$), então $g\rightarrow 2$. Como $y'\sim g\cos (t)$, então $y^2 + y'^2 \sim g^2$, e o plano de fases são trajetórias que convergem para um círculo de raio $2$. A Fig. \ref{fig:pert_mme_4} mostra o plano de fases obtido resolvendo-se numericamente a Eq. \ref{eq_mme_1}, variando-se o valor de $a$. Em ambos os casos, as trajetórias convergem para um círculo de raio $2$.

%% file: distribuicoes.tex
\chapter{Funções Generalizadas}
\label{chapter:distribuicoes}

\section{Introdução}
\label{section:introduction}

Em diversas situações práticas, são necessárias formulações da Física ou da Matemática Aplicada com existência de descontinuidades. Alguns exemplos são as interfaces em escoamentos bifásicos e as ondas de choque em escoamentos compressíveis. Nestes casos, a formulação do problema com funções \textit{ordinárias} não é adequada. Uma \textit{Função Generalizada}, ou \textit{distribuição}, ou ainda \textit{funcional} é uma generalização do conceito de função permitindo a formulação matemática de certos problemas. 

Seja $\phi (x)$ uma função de classe $C^\infty$ e de suporte compacto, conhecida. Uma distribuição é o processo de se atribuir um número $N_g$ a $\phi (x)$. Usualmente, representa-se uma distribuição por:

\begin{equation}
N_g\left[\phi (x)\right]\,=\,\int_{-\infty}^{\infty}g(x)\phi (x)\mathrm{d} x
\label{definition}
\end{equation}

A integral da Eq. \ref{definition} e $g(x)$ não possuem significado independente: elas são definidas pelo número $N_g\left[\phi (x)\right]$. A função generalizada também é comumente representada por

\begin{equation}
(g,\phi)
\label{definition2}
\end{equation}

\section{Função delta de Dirac}
\label{section:delta}

A função delta é definida por:

\begin{equation}
\int_{-\infty}^{\infty}\delta(x)\mathrm{d} x\,=\,1
\label{def_delta1}
\end{equation}

\noindent com

\begin{equation*}
\begin{array}{cc} \delta(x)=0 & x\neq 0 \end{array}
\end{equation*}

Para uma função $\phi (x)$ adequada, conhecida como \textit{função teste} (definida na seção \ref{section:espaco_teste}), a função delta possui a seguinte propriedade:  

\begin{equation}
\int_{-\infty}^{\infty}\delta(x)\phi (x)\mathrm{d} x\,=\,\phi (0)
\label{def_delta2}
\end{equation}     

\section{Significado Físico}
\label{section:sf}

\subsection{Percepção humana do Universo}
Considere que o valor $N_g\left[\phi (x)\right]$ representa uma observação (por exemplo uma força, ou uma diferença de potencial) cujo comportamento é normalmente associado a uma função. Uma função generalizada engloba as funções ordinárias, assim chamemos de $g(x)$ a função. Entretanto, nas medidas físicas (nossa percepção do Universo) nunca medimos a função $g(x)$, mas sim um número que é o resultado do experimento, isto é, o funcional $N_g\left[\phi (x)\right]$.

Desta forma, embora os comportamentos físicos sejam modelados por funções $g(x)$, nunca medimos tais funções. O que medimos são as respostas de um sistema $N_g\left[\phi (x)\right]$ em um dado meio $\phi (x)$. Assim, faz todo o sentido tratarmos nossas medidas físicas no sentido das distribuições, i.e., Eq. \ref{definition}.

\subsection{Singularidades do Universo}
No Universo e em alguns de nossos modelos matemáticos existem singularidades. Estas são, em geral, pontos, curvas ou superfícies
descontínuas (no espaço-tempo). Para uma região contendo uma dessas singularidades, uma função ordinária não pode ser definida. Vejamos um exemplo.

Seja um ponto material de massa unitária, isto é, a massa está contida em um volume nulo. Consideremos ainda que este ponto esteja em $x=0$. Se tentarmos calcular sua densidade no sentido das funções ordinárias, veremos que estamos diante de uma singularidade pois a densidade será infinita. Devemos então lançar mão das distribuições e definir a densidade material como $\delta(x)$ na Eq. \ref{def_delta1}. Para um ponto material de massa $m$ qualquer, a densidade é definida de acordo com

\begin{equation*}
\int_{-\infty}^{\infty}m\delta(x)\phi (x)\mathrm{d} x\,=\,m\phi (0)
\end{equation*}

\noindent onde $m\delta(x)$ é a densidade do ponto material em $x=0$.

\section{Espaço das funções teste}
\label{section:espaco_teste}

Vimos que o conceito de função generalizada depende de funções contínuas conhecidas como função teste. Desta forma, uma função generalizada é um funcional contínuo sobre um espaço de funções teste. Veremos agora qual é o espaço das funções teste.

Definiremos inicialmente o espaço das funções teste como o espaço de todas as funções $C^\infty$, e de suporte compacto em $\mathbb{R}^n$, denotado por $\mathcal{D}=\mathcal{D}(\mathbb{R}^n)$. Define-se então um espaço linear onde a operação de diferenciação em $\mathcal{D}$ é contínua em $\mathcal{D}$.

\begin{definition}
Seja a sequência $\phi_1(x),\,\phi_2(x),\,...\,\in\,\mathcal{D}$. Se:\\
(i) $\exists\;R>0\;|\;supp\,\phi (x)\,\subset\,U_R$\\
(ii) Para $(\beta_1,\,\beta_2,\,...,\,\beta_n)$ e $x\,\in\,\mathbb{R}^n$, tem-se que $D_\beta\phi_j(x)\,\rightarrow\,D_\beta\phi(x)$, $j\rightarrow\infty$\\
então diz-se que
\begin{center}
	$\phi_j(x)\,\rightarrow\,\phi(x)$, $j\rightarrow\infty$
\end{center}	
\end{definition}

O conjunto de funções teste cujos suportes estão contidos em uma região $\Omega\,\subset\,\mathbb{R}^n$ terão suporte $\mathcal{D}(\Omega)$.

Observação: a mudança não singular de variáveis e a multiplicação por uma função $a\,\in\,C^\infty(\mathbb{R}^n)$, são contínuas de $\mathcal{D}$ em $\mathcal{D}$.

Existem diversas possibilidades de funções teste. A forma canônica de função teste é chamada de \textit{função chapéu}:

\begin{equation}
\omega_\epsilon (x)\,=\,\left\{ \begin{array}{c} C_\epsilon\exp\left( -\frac{\epsilon^2}{\epsilon^2-x^2}\right)\; se\; \left| x\right|\leq\epsilon\\ \, \\ 0\; se\; \left| x\right|>\epsilon \\ \end{array}\right.
\label{funcao_chapeu}
\end{equation}

A constante $C_\epsilon$ é dada por

\begin{equation*}
\int_{-\infty}^{\infty}\omega_\epsilon (x)\mathrm{d} x\,=\,1
\end{equation*}

\begin{lemma}
\label{lemma_1}
Seja uma dada região $\Omega\,\in\,\mathbb{R}^n$ e um número $\epsilon>0$, $\exists\;\phi\;\in\;C^\infty(\mathbb{R}^n)\; |$\\
	\begin{center}
	$0\,\leq\,\phi(x)\,\leq\,1$\\
	$\phi (x)\,=\,1$, $x\,\in\,\Omega_\epsilon$\\
	$\phi (x)\,=\,0$, $x\,\notin \,\Omega_{3\epsilon}$\\
		\end{center}
onde tais regiões são contidas umas nas outras.
\end{lemma}


\begin{prova}
Deixado como exercício para o leitor.
\end{prova}

\begin{lemma}
Seja um número finito de vizinhanças $U(x_j;r_j),\;j\,=\,(1,\,...\,N)$, então $\exists\;\phi \eta_j\;\in\;\mathcal{D}\; |$\\
	\begin{center}
	$\phi (x)\,=\,\displaystyle\sum\limits_{j=1}^N {\phi(x)\eta_j(x)}$\\
	$supp\,\phi\eta_j\;\subset\;U(x_j;r_j)$
	\end{center}
isto é, o suporte de uma função teste, $\mathcal{D}(\Omega)$, pode ser coberto por um número finito de vizinhanças. 
\end{lemma}

\begin{prova}
Deixado como exercício para o leitor.
\end{prova}

\subsection{Suporte das funções teste}

Do Lemma \ref{lemma_1}, segue que:

\noindent se a região $\Omega$ for limitada, $\exists\;\phi\;\in\;\mathcal{D}\;|\;\phi (x)=1,\;x\;\in\;\Omega_{\epsilon}$;

\noindent o suporte $supp\,\phi (x)\;=\;\mathcal{D}(\Omega)$ é a região fechada para a qual $\phi (x)\,\neq\,0$ ($\phi$ tem suporte compacto).

\section{Espaço das distribuições}
\label{section:espaco_dist}

Uma função generalizada (no sentido de Sobolev-Schwartz) é um funcional linear sobre $\mathcal{D}$, e seu espaço é denotado por $\mathcal{D}'$. A ação de uma distribuição $g\,\in\,\mathcal{D}'$ sobre uma função teste $\phi (x)\,\in\,\mathcal{D}$ é representada pelas Equações \ref{definition} e \ref{definition2}. Usualmente se escreve $g(x)$ onde $x$ é o argumento das funções teste sobre as quais $g$ atua.

\begin{definition}
$(g,\phi)$ significa que um número complexo está associado a $\phi (x)\,\in\,\mathcal{D}$.
\end{definition}

\begin{definition}
$g$ é uma aplicação linear sobre $\mathcal{D}$. Para $\phi,\,\xi\;\in\;\mathcal{D}$ e $a,\,b\;\in\;\mathbb{C}$:
	\begin{center}
	$(g, a\phi+b\xi)\,=\,a(g,\phi)\,+\,b(g,\xi)$
	\end{center}
\end{definition}

\begin{definition}
$g$ é um funcional contínuo sobre $\mathcal{D}$. Para $\phi_j(x)\,\rightarrow\,\phi(x)$, $j\rightarrow\infty$:
	\begin{center}
	$(g,\phi_j)\;\rightarrow\;(g,\phi)\,,\;j\;\rightarrow\;\infty$
	\end{center}
\end{definition}

\begin{definition}
O espaço $\mathcal{D}'$ é um espaço linear sobre o corpo dos complexos. Para $g,\,u\;\in\;\mathcal{D}'$ $\phi,\,\xi\;\in\;\mathcal{D}$, $a,\,b,\,c,\,d\;\in\;\mathbb{C}$ e $\phi_j(x)\,\rightarrow\,\phi(x)$, $j\rightarrow\infty$:
	\begin{center}
	$(ag+bu, \phi)\,=\,a(g,\phi)\,+\,b(u,\phi)$\\
	$(ag+bu, c\phi+d\xi)\,=\,c(ag+bu,\phi)\,+\,d(ag+bu,\xi)$\\
	$(ag+bu,\phi_j)\;\rightarrow\;(ag+bu,\phi)\,,\;j\;\rightarrow\;\infty$
	\end{center}
\end{definition}

\begin{definition}
\label{defin_conv_distr}
Seja a sequência de funções $g_1,\,g_2,\,...\,\in\;\mathcal{D}'$. Esta sequência é convergente para $g\;\in\;\mathcal{D}'$ se para $\forall\;\phi\;\in\;\mathcal{D}$
	\begin{center}
	$(g_j,\phi)\;\rightarrow\;(g,\phi),\;j\,\rightarrow\,\infty$
	\end{center}
\end{definition}

\begin{lemma}
\label{lemma_complt}
O espaço $\mathcal{D}'$ é completo se dado $g_1,\,g_2,\,...\,\in\;\mathcal{D}'$ para $\forall\;\phi\;\in\;\mathcal{D}$
	\begin{center}
	$\displaystyle\lim_{j\to\infty}(g_j,\phi)\,=\,(g,\phi)$
	\end{center}
e $g\;\in\;\mathcal{D}'$ é única.
\end{lemma}

\begin{prova}
Ver \cite{Vladimirov}.
\end{prova}

\subsection{Suporte de uma distribuição}

\begin{definition}
$g\;\in\;\mathcal{D}'$ é nula em um aberto conexo $\Omega\;\subset\;\mathbb{R}^n$ se $\forall\;\phi\;\in\;\mathcal{D}(\Omega),\;(g,\phi)=0$. Diz-se então que $g(x)=0,\;x\;\in\;\mathcal{D}(\Omega)$.
\end{definition}

\begin{definition}
Diz-se que $g,\,u\;\;\in\;\mathcal{D}'$ são iguais em uma região $\Omega$ se $g-u=0,\;x\;\in\;\mathcal{D}(\Omega)$ Escreve-se que $g=u,\;x\;\in\;\mathcal{D}(\Omega)$.
\end{definition}

\begin{definition}
\label{defin_c_infinito}
$g\;\in\;\mathcal{D}'$ é $C^p(\Omega)$ se para $g_0(x)$ de classe $C^p(\Omega)$ e para $\forall\;\phi\;\in\;\mathcal{D}(\Omega)$
	\begin{center}
	$(g,\phi)\,=\,\int g_0(x)\phi (x)\mathrm{d} x$
	\end{center}
\end{definition}

\begin{lemma}
\label{lemma_vizinh}
Seja uma família contável de vizinhanças $U(x_j,r_j),\;j=1,\,2,\,...$ que recobre o $\mathbb{R}^n$. Se em cada $U(x_j,r_j)$ a distribuição $g$ coincidir com $g_j$ então $g$ é definida univocamente por suas componentes locais. 
\end{lemma}

\begin{prova}
Deixado como exercício para o leitor.
\end{prova}

Uma consequência do Lemma \ref{lemma_vizinh} é que para que uma função generalizada seja nula em determinada região é necesário e suficiente que ela seja nula na vizinhança de cada um dos pontos da região.

\begin{corolary}
A união de todos os abertos onde $g=0$ é um conjunto aberto. Este é o maior conjunto onde $g$ é nula. Usualmente este conjunto é denotado por $Z_g$
\end{corolary}

\begin{prova}
Segue imediatamente do Lemma \ref{lemma_vizinh}.
\end{prova}

\begin{definition}
\label{defin_supp_g}
O suporte de g, $supp\,g$, é o complemento de $Z_g$.
\end{definition}

\begin{corolary}
O $supp\,g$ é um conjunto fechado (em $\mathbb{R}^n$).
\end{corolary}

\begin{prova}
Segue imediatamente do Lemma \ref{lemma_vizinh} e da Definição \ref{defin_supp_g}.
\end{prova}

\begin{corolary}
Se $supp\,g$ é um conjunto limitado, então $g$ é de suporte compacto.
\end{corolary}

\begin{prova}
Todo conjunto limitado e fechado no $\mathbb{R}^n$ é compacto.
\end{prova}

\section{Distribuições regulares e singulares}
\label{section:reg_sing}

\begin{definition}
Distribuições Regulares são as funções generalizadas que podem ser definidas em termos de funções localmente integráveis em $\mathbb{R}^n$ (funções ordinárias). Todas as outras distribuições são ditas singulares.
\end{definition}

\begin{lemma}
(Du Bois Reymond) Para que uma função $f(x)$ localmente integrável em uma região $\Omega$ seja nula nesta região no sentido das distribuições, é necessário e suficiente que $f(x)=0$ em quase todo o lugar de $\Omega$.
\end{lemma}

\begin{prova}
Segue do Lemma \ref{lemma_vizinh}.
\end{prova}

\begin{corolary}
Uma distribuição regular é definida por uma única função localmente integrável. Cada função localmente integrável no $\mathbb{R}^n$ pode ser identificada como uma distribuição:
	\begin{center}
	$(f,\phi)\,=\,\displaystyle\int_{-\infty}^{\infty}f(x)\phi (x)\mathrm{d} x$
	\end{center}
\end{corolary}

\begin{prova}
Segue do Lemma de Du Bois Reymond.
\end{prova}

\begin{corolary}
Seja uma sequência de funções localmente integráveis $f_j(x),\;j=1,\,2,\,...\;\in\;\mathbb{R}^n$. Se ela converge uniformemente, sobre qualquer compacto, para $f(x)$, então ela converge para $f(x)$ em $\mathcal{D}'(\mathbb{R}^n)$.
	\begin{center}
	$(f_j,\phi)\,=\,\displaystyle\int_{-\infty}^{\infty}f_j(x)\phi (x)\mathrm{d} x\;\rightarrow\;(f,\phi)\,=\,\displaystyle\int_{-\infty}^{\infty}f(x)\phi (x)\mathrm{d} x,\;j\,\rightarrow\,\infty$
	\end{center}
\end{corolary}

\begin{prova}
Segue do Lemma \ref{lemma_complt}.
\end{prova}

\begin{lemma}
Lemma de Riemann-Lebesgue. Seja $\phi(x)$ uma função limitada e com derivada limitada em $[a,b]\,\subset\,\mathbb{R}$. Para esta função, vale:
	\begin{center}
	$\displaystyle\lim_{k\to\infty}\int_{a}^{b}e^{-ikx}\phi (x)\mathrm{d} x\,=\,0$
	\end{center}
\end{lemma}

\begin{prova}
$\int_{a}^{b}e^{-ikx}\phi (x)\mathrm{d} x\,=\,\frac{1}{ik}\left[ \phi (a)e^{-ika}-\phi (b)e^{-ikb}\right]\,+\,\frac{1}{ik}\int_{a}^{b}e^{-ikx}\phi '(x)\mathrm{d} x\,$

E, como $\phi (x)$ e $\phi '(x)$ são limitadas no intervalo considerado, o limite quando $k\,\rightarrow\,\infty$ é nulo.
\end{prova}

\begin{definition}
É impossível identificar uma distribuição singular com qualquer função localmente integrável (função ordinária).
\end{definition}

\begin{prop}
A função delta de Dirac $\delta(x)$ é uma distribuição singular.
\end{prop}

\begin{prova}
Deixado a cargo do leitor.
\end{prova}

\section{Operações com distribuições}
\label{section:propriedades}

Nesta seção são apresentadas algumas operações matemáticas no sentido das distribuições. Apenas algumas demostrações são feitas aqui, para as demais sugere-se que o leitor consulte as seguintes obras \cite{Schwartz_3}, \cite{Farassat} e \cite{Vladimirov}.

\subsection{Soma de distribuições}
A soma de duas distribuições $g(x)=g_1(x)+g_2(x)$ é dada por:

\begin{equation}
\int_{-\infty}^{\infty}g(x)\phi (x)\mathrm{d} x\, = \, \int_{-\infty}^{\infty}g_1(x)\phi (x)\mathrm{d} x\,+\,\int_{-\infty}^{\infty}g_2(x)\phi (x)\mathrm{d} x
\end{equation}

\subsection{Linearidade}
Para $g,\,u\;\in\;\mathcal{D}'$ $\phi,\,\xi\;\in\;\mathcal{D}$, $a,\,b,\,c,\,d\;\in\;\mathbb{C}$ e $\phi_j(x)\,\rightarrow\,\phi(x)$, $j\rightarrow\infty$:

\begin{equation}
(ag+bu, \phi)\,=\,a(g,\phi)\,+\,b(u,\phi)
\end{equation}

\begin{equation}
(ag+bu, c\phi+d\xi)\,=\,c(ag+bu,\phi)\,+\,d(ag+bu,\xi)
\end{equation}

\subsection{Produto de uma distribuição por uma função ordinária}
Seja $f(x)$ uma função localmente integrável. Então:

\begin{equation}
\int_{-\infty}^{\infty}\left(g(x)f(x)\right)\phi (x)\mathrm{d} x\, = \, \int_{-\infty}^{\infty}g(x)\left(f(x)\phi (x)\right)\mathrm{d} x
\label{eq_prod_distr_funcao}
\end{equation}

\subsection{Mudança de Variáveis}
Para $a,\,x_0\;\in\;\mathbb{C}$ 

\begin{equation}
\int_{-\infty}^{\infty}g(ax)\phi (x)\mathrm{d} x\,=\,\frac{1}{\left| a\right|}\int_{-\infty}^{\infty}g(x)\phi (x/a)\mathrm{d} x
\end{equation}

\begin{equation}
\int_{-\infty}^{\infty}g(x-x_0)\phi (x)\mathrm{d} x\,=\,\int_{-\infty}^{\infty}g(x)\phi (x+x_0)\mathrm{d} x
\label{eq_transl}
\end{equation}

A Eq. \ref{eq_transl} define uma operação de translação.

\subsection{Convolução}

\begin{equation}
\int_{-\infty}^{\infty} \left[\int_{-\infty}^{\infty}g_1(\xi) g_2(x-\xi )\mathrm{d} \xi\right]\phi (x)\mathrm{d} x\,=\,\int_{-\infty}^{\infty} g_1(\xi) \left[\int_{-\infty}^{\infty}g_2(x-\xi )\phi (x)\mathrm{d} x\right]\mathrm{d} \xi
\end{equation}

\subsection{Derivação}

\begin{equation}
\frac{d^n N_g}{dx^n}\,=\,\int_{-\infty}^{\infty}\frac{d^ng(x)}{dx^n}\phi (x)\mathrm{d} x\,=\left( -1\right)^n\,\int_{-\infty}^{\infty}g(x)\frac{d^n\phi (x)}{dx^n}\mathrm{d} x
\label{eq_derivada}
\end{equation}

A primeira passagem mostra que, quando a derivada é tomada no sentido das distribuições, a ordem das operações de derivação e de integração podem ser alternadas. A passagem do segundo para o terceiro membro da equação é obtida diretamente de uma integração por partes, lembrando que o suporte das funções teste é compacto.

\begin{corolary}
Qualquer distribuição é infinitamente diferenciável.
\end{corolary}

\begin{prova}
Segue imediatamente do fato que $\phi\;\in\;\mathcal{D}$ é $C^{\infty}(\Omega)$ e da Eq. \ref{eq_derivada}.
\end{prova}

\begin{corolary}
A ordem da derivação não altera o resultado.
\end{corolary}

\begin{prova}
Ver \cite{Vladimirov}.
\end{prova}

\begin{corolary}
A Regra da Cadeia vale para derivação de uma distribuição $g\;\in\;\mathcal{D}'$ por uma função $a (x)\;\in\;C^\infty(\mathbb{R}^n)$:

\begin{equation}
\frac{\partial ag}{\partial x_i}\,=\,g\frac{\partial a}{\partial x_i}\,+\,a\frac{\partial g}{\partial x_i},\;1=1,2,...,n
\label{eq_cadeia}
\end{equation}

\end{corolary}

\begin{prova}
$\partial_x(ga,\phi)\,=\,(\partial_x(ga),\phi)\,=\,-(ga,\partial_x(\phi))\,=\,-(g,a\partial_x(\phi))\,=\,-(g,\partial_x(a\phi)-\phi\partial_x(a))\,=\,-(g,\partial_x(a\phi))+(g,\phi\partial_x(a))\,=\,(\partial_xg,(a\phi))+(g\partial_x(a),\phi)\,=\,(a\partial_xg+g\partial_xa,(\phi))$.
\end{prova}

\subsection{Transformada de Fourier}
A seguinte propriedade é válida para transformadas de Fourier de distribuições:

\begin{equation}
\int_{-\infty}^{\infty}\hat{g}(k)\phi (x)\mathrm{d} x\,=\,\int_{-\infty}^{\infty}g(x)\hat{\phi} (k)\mathrm{d} x
\label{eq_fourier_gen}
\end{equation}
 
\noindent onde o símbolo $\hat{ }$ representa a transformada de Fourier.

\section{Operações matemáticas com as funções delta e de Heaviside}

\textbf{Função de Heaviside}

A função de Heaviside $h(x)$, também conhecida como função degrau, é definida como:

\begin{equation}
h(x)\,=\,\left\{ \begin{array}{c} 1\; se\; x>0\\ \, \\ 0\; se\; x<0 \\ \end{array}\right.
\label{funcao_heaviside}
\end{equation}

\noindent o que pode ser denotado como:

\begin{equation}
H\left[\phi\right]\,=\,\int_{-\infty}^{\infty}h(x)\phi (x)\mathrm{d} x\,=\,\int_{0}^{\infty}\phi (x)\mathrm{d} x
\label{funcao_heaviside2}
\end{equation}

Como veremos no que segue, ela está diretamente relacionada à função delta de Dirac e possui diversas aplicações em Física e Matemática Aplicada.

\textbf{Multiplicação de $\delta$ por uma função ordinária}

Seja $a(x)\;\in\;\mathcal{D}'$ uma função localmente integrável. Das Eqs. \ref {def_delta2} e \ref{eq_prod_distr_funcao}, temos que:

\begin{equation}
a\delta[\phi]\,=\,\delta[a\phi]\,=\,a(0)\phi(0)
\end{equation}

\noindent que também é escrita simbolicamente como:

\begin{equation}
a(x)\delta (x)=a(0)\delta(0)
\end{equation}

\noindent OBS: cuidado com este tipo de notação, que causa certa confusão. Note que o seu lado esquerdo representa o que estaria sendo multiplicado por $\phi$ dentro da integral.

\textbf{Translação de $\delta$}

Seja $x_0$ uma dada posição da variável $x$. Da Eq. \ref{eq_transl}:

\begin{equation}
\delta\left[\phi(x+x_0)\right]\,=\,\int_{-\infty}^{\infty}\delta(x)\phi (x+x_0)\mathrm{d} x\,=\,\phi (x_0)
\end{equation}

\textbf{sequências tendendo a $\delta$}

Da Definição \ref{defin_conv_distr} e do Lemma \ref{lemma_complt}, se a sequência $g_j(x)$ tende à $\delta(x)$, então:

\begin{equation}
\displaystyle\lim_{j\to\infty}(g_j,\phi)\,=\,(\delta,\phi)\,=\,\phi (0)
\end{equation}

Por exemplo, seja a sequência:

\begin{equation}
g_j(x)\,=\,\left\{ \begin{array}{c} j^2\left(\frac{1}{j}-\left| x\right|\right)\; se\; \left| x\right|\leq \frac{1}{j}\\ \, \\ 0\; se\; \left| x\right| > \frac{1}{j} \\ \end{array}\right.
\end{equation}

\noindent e como $\displaystyle\lim_{j\to\infty}(g_j,\phi)\,=\,(\delta,\phi)$, então $\displaystyle\lim_{j\to\infty}(g_j,\phi)\,=\,\phi (0)$.

\textbf{Derivada de $\delta$}

Da Eq. \ref{eq_derivada}:

\begin{equation}
\delta '[\phi]\,=\,\int_{-\infty}^{\infty}\delta '(x)\phi (x)\mathrm{d} x\,=\,-\int_{-\infty}^{\infty}\delta(x)\phi '(x)\mathrm{d} x\,=\,-\phi '(0)
\end{equation}

\noindent onde o ' significa derivada em relação à variável independente da função teste (no caso em questão, $x$).

\textbf{Derivada de $h$}

Das Eqs. \ref{eq_derivada} e \ref{funcao_heaviside2}, e lembrando que $supp\,\phi$ é compacto:

\begin{equation}
H'\left[\phi\right]\,=-\,\int_{0}^{\infty}\phi '(x)\mathrm{d} x\,=\,-\phi (\infty)\,+\,\phi (0)\,=\,\phi (0)\,=\,\delta\left[\phi\right]
\end{equation}

\noindent isto é,

\begin{equation}
h'(x)\,=\,\delta (x)
\end{equation}

\textbf{Transformada de Fourier de $\delta$}

Seja a transformada de Fourier de uma função $\phi (x)$

\begin{equation}
\int_{-\infty}^{\infty} \phi (x)e^{2\pi ikx}\mathrm{d} x
\end{equation}

Da Eq. \ref{eq_fourier_gen}:

\begin{equation}
\hat{\delta}[\phi]\,=\,\delta[\hat{\phi}]\,=\,\hat{\phi} (0)\,=\,\int_{-\infty}^{\infty} \phi (x)\mathrm{d} x
\end{equation}

\noindent logo, da passagem do terceiro para o quarto termo,

\begin{equation}
\hat{\delta}(k)\,=\,1
\end{equation}

\textbf{Derivadas de funções descontínuas}

Seja $f(x)$ uma função contínua por partes, com uma descontinuidade (em uma dimensão) em $x=x_0$. Esta descontinuidade pode ser interpretada como um salto em $x_0$:

\begin{equation}
\Delta f(x_0)\,=\,f(x_0^+)\,-\,f(x_0^-)
\end{equation}

A derivada de uma função deste tipo pode ser calculada no sentido das distribuições. Seja $\phi\,\in\,\mathcal{D}$ e $x_0\,\in\,supp\,\phi$. Considere um funcional $G[\phi (x)]$ relacionado a $f(x)$ e que $supp\,\phi\,=\,[a,b]$. Então, para $\epsilon\,=\,x_0^+-x_0^-\,\rightarrow\,0$:

\begin{equation*}
G'[\phi ]\,=\,-G[\phi ']\,=\,-\int_{-\infty}^{+\infty}f(x)\phi '(x)\mathrm{d} x\,=\,-\int_{-\infty}^{x_0^-}f(x)\phi '(x)\mathrm{d} x -\int_{x_0^+}^{+\infty}f(x)\phi '(x)\mathrm{d} x 
\end{equation*}

\begin{equation*}
\,=\, -f(x)\phi (x)|_{-\infty}^{x_0^-} + \int_{-\infty}^{x_0^-}f'(x)\phi (x)\mathrm{d} x  -f(x)\phi (x)|_{x_0^+}^{+\infty} + \int_{x_0^+}^{+\infty}f'(x) \phi (x)\mathrm{d} x 
\end{equation*}

\begin{equation*}
\,=\, -f(x_0^-)\phi (x_0^-)+ f(x_0^+)\phi (x_0^+) + \int_{-\infty}^{+\infty}f'(x)\phi (x)\mathrm{d} x 
\end{equation*}

\noindent logo,

\begin{equation}
G'[\phi ]\,=\,\int_{a}^{b}f'(x)\phi (x)\mathrm{d} x \,+\,\Delta f(x_0)\phi (x_0)
\end{equation}


Onde:

\begin{equation*}
\phi (x_0)\,=\,\delta\left[ \phi(x+x_0)\right]
\end{equation*}

\noindent e assim,

\begin{equation}
G'[\phi ]\,=\,\int_{a}^{b}f'(x)\phi (x)\mathrm{d} x \,+\,\Delta f(x_0)\delta\left[ \phi(x+x_0)\right]
\end{equation}

Usualmente escrevemos:

\begin{equation}
g'(x)\,=\,f'(x)\,+\,\Delta f(x_0)\delta (x-x_0)
\label{eq_deriv_1d}
\end{equation}

Note que, para um intervalo $[a,b]$ qualquer, a derivada de uma função com salto no sentido das distribuições, $g'(x)$, é diferente da derivada no sentido ``localmente integrável", $f'(x)$. A derivada no sentido das distribuições leva em conta a derivada da parte contínua, assim como do salto $\Delta f$.

Note ainda que a integração da Eq. \ref{eq_deriv_1d} recupera a função descontínua original:

\begin{equation}
\int_{a}^{b}g'(x)\mathrm{d} x\,=\int_{a}^{b}f'(x)\mathrm{d} x \,+\,\Delta f(x_0)h(x-x_0)
\end{equation}

\noindent onde aparece claramente a descontinuidade do tipo degrau (via a função de Heaviside).

\subsection{Regra de Leibniz de diferenciação de integral}
Considere a variável independente $\xi$, a função contínua $\Psi\,\in\,\mathcal{D}$ e os limites $A(\xi)$ e $B(\xi)>A(\xi)$. Deseja-se calcular a seguinte derivada:

\begin{equation}
E(\xi)\,=\,\frac{d}{d\xi}\int_{A(\xi)}^{B(\xi)}\Psi (x,\xi)\mathrm{d}x
\label{eq_leibniz_1}
\end{equation}

Podemos definir uma função generalizada $T$ com o auxílio da função de Heaviside:

\begin{equation*}
T(x,\xi)\,=\,h(x-A(\xi))h(B(\xi)-x)
\end{equation*}

\noindent de forma que $T(x,\xi)=1$ em $]A,B[$ e nula em $]-\infty, A[$ e $]B,+\infty[$. Desta maneira, a Eq. \ref{eq_leibniz_1} pode ser escrita como:

\begin{equation}
E(\xi)\,=\,\frac{d}{d\xi}\int_{-\infty}^{+\infty}T(x,\xi)\Psi (x,\xi)\mathrm{d}x\,=\,\int_{-\infty}^{+\infty}\frac{\partial T(x,\xi)}{\partial\xi}\Psi (x,\xi)+\frac{\partial \Psi (x,\xi)}{\partial\xi}T(x,\xi)\mathrm{d}x
\label{eq_leibniz_2}
\end{equation}

\noindent onde:

\begin{equation*}
\frac{\partial T(x,\xi)}{\partial\xi}\,=\,-A'(\xi)\delta(x-A(\xi))h(B(\xi)-x)+B'(\xi)h(x-A(\xi))\delta(B(\xi)-x)
\end{equation*}

\noindent Note entretanto que $\delta(x-A(\xi))h(B(\xi)-x)\,=\,\delta(x-A(\xi))$ e que $h(x-A(\xi))\delta(B(\xi)-x)\,=\,\delta(B(\xi)-x)$ pois $B(\xi)>A(\xi)$ logo $h=1$. Assim:

\begin{equation}
\frac{\partial T(x,\xi)}{\partial\xi}\,=\,-A'(\xi)\delta(x-A(\xi))+B'(\xi)\delta(B(\xi)-x)
\label{eq_leibniz_3}
\end{equation}

\noindent Inserindo a Eq. \ref{eq_leibniz_3} na Eq. \ref{eq_leibniz_2}:
\begin{equation}
E(\xi)\,=\,\int_{A(\xi)}^{B(\xi)}\frac{\partial \Psi (x,\xi)}{\partial\xi}\mathrm{d}x\,+\,B'(\xi)\Psi (B(\xi),\xi)\,-\,A'(\xi)\Psi (A(\xi),\xi)
\label{eq_leibniz_4}
\end{equation}

\noindent que é o Teorema de Leibniz para diferenciação da integral.

\section{Distribuições no espaço multi-dimensional}

\subsection{Função delta}
No espaço tridimensional, a função delta de Dirac é definida como:

\begin{equation}
\int_{-\infty}^{\infty}\delta(\vec{x})\phi (\vec{x})\mathrm{d} \vec{x}\,=\,\phi (0)
\end{equation}

\subsection{Distribuição de camada simples}

A função $\delta_s$ ou $\delta(\Sigma)$ é uma generalização da função $\delta$ ``pontual".

\begin{definition}
\label{def_camada_simples}
Seja $S$ uma superfície contínua por partes no $\mathbb{R}^n$ e seja $\mu (x)$ uma função contínua sobre $S$. A distribuição de camada simples sobre a superfície $S$, para $\forall\,\phi\,\in\,\mathcal{D}$ é definida:

\begin {equation}
(\mu\delta_s,\phi)\,=\,\int_{-\infty}^{\infty}\ \mu (x)\delta(\Sigma )\phi (\vec{x})\mathrm{d} \vec{x}\,=\,\int_{S}\mu (x)\phi (x)\mathrm{d} S
\end{equation}

\noindent onde $\Sigma$ define uma superfície. Note que $\mu\delta_s\,\in\,\mathcal{D}'$ e que $\mu\delta_s=0$ para $x\,\notin\,S$ de forma que $\mu\delta_s\,\subset\,S$.
\end{definition}

A Definição \ref{def_camada_simples} mostra que o funcional $\delta_s[\phi]$ só não é nulo nos pontos do espaço pertencentes à superfície  $\Sigma$.

\subsection{Divergente, gradiente e rotacional}

Em espaços multidimensionais, a Eq. \ref{eq_deriv_1d} pode ser facilmente generalizada. Isto nos fornece então, no sentido das distribuições, o gradiente (Eq. \ref{eq_grad}), o divergente (Eq. \ref{eq_div}) e o rotacional (Eq. \ref{eq_rot}) de funções descontínuas.

\begin{equation}
\nabla g\,=\,\nabla f\,+\,\Delta f\nabla (\Sigma)\delta_s
\label{eq_grad}
\end{equation}

\begin{equation}
\nabla\cdot\vec{g}\,=\,\nabla\cdot\vec{f}\,+\,\Delta\vec{f}\cdot\nabla (\Sigma)\delta_s
\label{eq_div}
\end{equation}

\begin{equation}
\nabla\times\vec{g}\,=\,\nabla\times\vec{f}\,+\,\Delta\vec{f}\times\nabla (\Sigma)\delta_s
\label{eq_rot}
\end{equation}

\subsection{Teorema da divergência de Gauss}

Seja $\Omega\,\subset\,\mathbb{R}^3$ uma variedade tridimensional simplesmente conexa, cuja fronteira é a superfície $\partial\Omega$. Considere o seguinte campo vetorial descontínuo:

\begin{equation}
\vec{\psi}(\vec{x})\,=\,\left\{ \begin{array}{c} \vec{\phi}_g(\vec{x})\; se\; \vec{x}\,\in\,\Omega \\ \, \\ 0\; se\; \vec{x}\,\notin\,\Omega \\ \end{array}\right.
\end{equation}

Seja uma distribuição $\vec{\psi}_g(\vec{x})$ associada ao campo $\vec{\psi}(\vec{x})$. O divergente desta distribuição é dado pela Eq. \ref{eq_div}:

\begin{equation}
\nabla\cdot\vec{\psi}_g\,=\,\nabla\cdot\vec{\psi}\,+\,\Delta\vec{\psi}\cdot\nabla (\partial\Omega)\delta_s
\end{equation}

\noindent e, observando que $\Delta\vec{\psi}\,=\,-\vec{\phi_g}$, que $\nabla\cdot\vec{\psi}\,=\,\nabla\cdot\vec{\phi_g}$ e que $\nabla (\partial\Omega)\,=\,\vec{n}$ (apontando para fora):

\begin{equation}
\nabla\cdot\vec{\psi}_g\,=\,\nabla\cdot\vec{\phi}_g\,-\,\vec{\phi}_g\cdot\vec{n}\delta_s
\label{eq_gauss_a}
\end{equation}

A integral em todo $\mathbb{R}^3$ do membro esquerdo da Eq. \ref{eq_gauss_a} fornece:

\begin{equation}
\int_{\mathbb{R}^3}\nabla\cdot\vec{\psi}_g\mathrm{d}\vec{x}\,=\,\,\int_{\mathbb{R}^3} \frac{\partial\psi_{gi}}{\partial x_i} \mathrm{d}x_j\,=\,\int_{\mathbb{R}^2} \left.\psi_{gi}\right|_{-\infty}^{+\infty}\left.dx_j\right|_{j\neq i}\,=\,0
\end{equation}

\noindent logo,

\begin{equation}
\int_{-\infty}^{\infty}\nabla\cdot\vec{\phi}_g\mathrm{d}\vec{x}\,=\,\int_{-\infty}^{\infty}\vec{\phi}_g\cdot\vec{n}\delta_s\mathrm{d}\vec{x}
\end{equation}

\noindent e assim, para $\phi_n\,=\,\vec{\phi}_g\cdot\vec{n}$:

\begin{equation}
\int_{\Omega}\nabla\cdot\vec{\phi}_g\mathrm{d}\vec{x}\,=\,\int_{\partial\Omega}\phi_n\mathrm{d}S
\label{eq_gauss_b}
\end{equation}

Agora, suponha que exista uma descontinuidade dentro da região $\Omega$. Seja uma superfície $\Sigma\,\subset\,\Omega$ onde ocorre um salto em $\vec{\phi}_g$. Neste caso, $\vec{\phi}_g$ pode ser considerada como uma distribuição, de forma que, ao atravessar a superfície $\Sigma$, a descontinuidade deve ser considerada. Seja $\vec{\phi(x)}$ uma função do campo vetorial atravessando $\Sigma$, de forma a haver uma descontinuidade. Para o operador divergente existente na Eq. \ref{eq_gauss_b}:

\begin{equation}
\nabla\cdot\vec{\phi}_g\,=\,\nabla\cdot\vec{\phi}\,+\,\Delta\vec{\phi}\cdot\nabla (\Sigma)\delta_s
\end{equation}

\noindent e, inserindo este divergente na Eq. \ref{eq_gauss_b}:

\begin{equation}
\int_{\Omega}\nabla\cdot\vec{\phi}\mathrm{d}\vec{x}\,=\,\int_{\partial\Omega}\phi_n\mathrm{d}S\,-\,\int_{\Sigma}\Delta\vec{\phi}\cdot \vec{n}_{\Sigma}\mathrm{d}S
\label{eq_gauss_c}
\end{equation}

\noindent onde $\Delta\vec{\phi}$ é o salto, isto é, a diferença entre os valores que a função $\vec{\phi(x)}$ possui em ambos os lados da superfície $\Sigma$ e $\nabla (\Sigma )\,=\,\vec{n}_{\Sigma}$.

\subsection{Derivada da função delta}

Para a derivada da função delta de camada simples, pode ser demonstrado que:

\begin{equation}
(\delta'_s,\phi)\,=\,\int_{-\infty}^{\infty}\ \delta'_s\phi (\vec{x})\mathrm{d} \vec{x}\,=\,\int_{S}-\frac{\partial\phi}{\partial n}+2\mathcal{H}_s(x)\phi(x)\mathrm{d} S
\end{equation}

\noindent onde $\mathcal{H}_s(x)$ é a curvatura média local da superfície.

\subsection{Laplaciano de uma função descontínua}

Seja $\Omega\,\subset\,\mathbb{R}^n$ uma região de fronteira $\partial\Omega$ (logo uma variedade de dimensão $n-1$). Seja $\Omega'$ o complemento de $\Omega$. Seja $f(x_i)\,\in\,\mathbb{C}$, $i=1,2,...,n$ uma função descontínua sobre a fronteira $\partial\Omega$. A descontinuidade pode ser escrita como: 

\begin{equation*}
\Delta f_s\,=\,\lim_{x_i\to x_s,\, x_i\,\in\,\Omega'}f(x_i)-\lim_{x_i\to x_s,\, x_i\,\in\,\Omega}f(x_i)
\end{equation*}

\noindent onde $x_s$ são as coordenadas da superfície $\partial\Omega$.

Pode-se provar que o laplaciano de $f$ no sentido das distribuições $\nabla^2_{gen}f$ é dado por:

\begin{equation}
\nabla^2_{gen}f\,=\,\nabla^2f\,+\,\frac{\partial (\Delta f_s\delta_s)}{\partial n}\,+\,\frac{\partial f}{\partial n}\delta_s
\label{eq_lapl_general_1}
\end{equation}

\noindent onde $\delta_s$ é a distribuição de camada simples que atua sobre a fronteira $\partial\Omega$ e $n$ é a normal à tal fronteira.

Ainda, se considerarmos o caso em que $supp\,f$ é tal que $f=0$ em $x\,\in\,\Omega'$:

\begin{equation}
\nabla^2_{gen}f\,=\,\nabla^2f\,-\,\frac{\partial (f\delta_s)}{\partial n}\,-\,\frac{\partial f}{\partial n}\delta_s
\label{eq_lapl_general_2}
\end{equation}

A prova da Eq. \ref{eq_lapl_general_1} é deixada como exercício.

\subsection{Fórmula de Green}

A fórmula de Green (clássica) pode ser facilmente provada utilizando-se o laplaciano no sentido das distribuições. Seja a função, contínua por partes, $f(x)$ e $supp\,f\,\subset\,\Omega$, de forma que $f=0$ em $x\,\in\,\Omega'$. Tomando-se o laplaciano generalizado de $f$:

\begin{equation*}
\nabla_{gen}^2(f,\phi)\,=\,(\nabla^2f,\phi)\,-\,\left(\frac{\partial (f\delta_s)}{\partial n},\phi\right) \,-\,\left(\frac{\partial f}{\partial n}\delta_s,\phi\right)
\end{equation*}

\begin{equation*}
=\,(\nabla^2f,\phi)\,+\,\left( f\delta_s,\frac{\partial\phi}{\partial n}\right) \,-\,\left(\frac{\partial f}{\partial n}\delta_s,\phi\right)
\end{equation*}

\noindent e, como $\phi$ é $C^\infty$, $(\nabla_{gen}^2f,\phi)\,=\,(f,\nabla^2\phi)$, logo:

\begin{equation}
\int_{\Omega}\left( f\,\nabla^2\phi -\nabla^2f\,\phi\right)\mathrm{d}\vec{x}=\,\int_{\partial\Omega}\left( f\frac{\partial\phi}{\partial n}-\frac{\partial f}{\partial n}\phi\right)\mathrm{d}S
\label{eq_form_green}
\end{equation}

\section{Aplicações}

\subsection{Função de Green e sua utilização na solução de equações}

\subsubsection{Função de Green em um espaço 3D infinito e isotrópico}

Seja um $\Omega\,\subset\,\mathbb{R}^3$ uma variedade tridimensional simplesmente conexa, de fronteira $\partial\Omega$. Nos interessaremos à solução da equação:

\begin{equation}
\nabla^2G(\vec{x},\vec{x_b})\,=\,\delta(\vec{x})
\end{equation}

\begin{equation}
\int_{\Omega}\delta(\vec{x})\mathrm{d} \forall\,=\,1
\end{equation}

\noindent nesta variedade, onde $\vec{x}$ e $\vec{x_b}$ são os vetores posição de pontos vizinhos. Consideremos que esta esta variedade possua propriedades isotrópicas. Neste caso $G$ depende apenas de $r\,=\,|\vec{x}-\vec{x_b}|$. Assim:

\begin{equation}
\nabla^2G(r)\,=\,\delta(\vec{x})
\end{equation}

Seja $\Omega_c\,\subset\,\Omega$ compreendido entre duas bolas, uma externa e uma interna, formando desta forma uma ``casca" na região em que	$\,\displaystyle\int_{\Omega_c}\delta(\vec{x})\mathrm{d} \forall\,=\,0$. Para esta região $\nabla^2G(r)\,=\,0$ e assim:

\begin{equation}
\int_{\Omega_c}\nabla^2G\mathrm{d} \forall\,=\,\int_{\partial\Omega_e\cup\partial\Omega_i}\vec{n}\cdot\nabla G\mathrm{d} S\,=\,\int_{\partial\Omega_e\cup\partial\Omega_i}\frac{\partial G}{\partial n} \mathrm{d} S\,=\,0
\end{equation}

\noindent onde $\partial\Omega_e\cup\partial\Omega_i$ é a fronteira da região $\Omega_c$. Da equação anterior segue que os fluxos através das superfícies interna $\partial\Omega_i$ e externa $\partial\Omega_e$ devem ser os mesmos, logo 

\begin{equation}
\int_{\partial\Omega}G' \mathrm{d} S\,=\,cte\,=k\;\Rightarrow\;G(r)\,=\,-\frac{k}{4\pi r}
\end{equation}

Considere agora $\Omega_i\,\subset\,\Omega$, simplesmente conexa e com mesmo centro (logo uma bola concêntrica dentro de $\Omega$), tal que nesta região $\,\displaystyle\int_{\Omega_i}\delta(\vec{x})\mathrm{d} \forall\,=\,1$. Nesta região

\begin{equation}
\int_{\Omega_i}\nabla^2G\,\mathrm{d} \forall\,=\,\int_{\Omega_i}\delta(\vec{x})\mathrm{d} \forall\,=\,1
\end{equation}

\noindent logo $k=1$ e a função de Green em $\Omega\,\subset\,\mathbb{R}^3$ é dada por

\begin{equation}
G(r)\,=\,-\frac{1}{4\pi r}
\label{eq_green_3d}
\end{equation}

\subsubsection{Utilização em equações}

Seja, por exemplo, o seguinte problema

\begin{equation}
\left\{ \begin{array}{c} u''(x)\,=\,f(x)\\ \, \\ u(0)\,=\,u(1)\,=0 \\ \end{array}\right.
\label{eq_ex_green_1}
\end{equation}

\noindent onde $f(x)$ é um dado do problema, mas para o qual propriedades de continuidade não são asseguradas. Desta forma, é preferível fazer a derivação da Eq. \ref{eq_ex_green_1} no sentido das distribuições:

\begin{equation}
\left\{ \begin{array}{c} D^2u(x)\,=\,f(x)\\ \, \\ u(0)\,=\,u(1)\,=0 \\ \end{array}\right.
\label{eq_ex_green_2}
\end{equation}

\noindent Na Eq. \ref{eq_ex_green_2} utiliza-se a notação $D$ para derivação no sentido da distribuições em relação à variável independente.

Vamos generalizar e chamar de $L$ o operador diferencial do problema. No caso em questão, $L=D^2$. No espaço de todos os operadores deste tipo:

\begin{equation}
Lu(x)\,=\,f(x)
\label{eq_ex_green_3}
\end{equation}

Vamos também considerar que exista o inverso de tal operador, isso é, $L^{-1}$, de tal forma que $u=L^{-1}f$. Faz-se aqui ainda uma hipótese: $L^{-1}$ é um operador do tipo \textit{Kernel}

\begin{equation}
L^{-1}v(x)\,\equiv\,\int g(x,t)v(t)dt
\end{equation}

\noindent onde $g$ é o núcleo (Kernel) do operador. Desta forma:

\begin{equation}
f(x)\,=\,Lu(x)\,=\,LL^{-1}f(x)\,=\,L\int g(x,t)f(t)dt\,=\,\int Lg(x,t)f(t)dt
\label{eq_ex_green_4}
\end{equation}

Da Eq. \ref{eq_ex_green_4}, percebe-se que $Lg(x,t)\,=\,\delta_x(t)$, logo no caso do espaço tridimensional, $g=G(r)$ dado pela Eq. \ref{eq_green_3d}.

Também da Eq. \ref{eq_ex_green_4}, obtém-se:

\begin{equation}
u(x)\,=\,\int g(x,t)f(t)dt
\label{eq_ex_green_5}
\end{equation}

\noindent solução da Eq. \ref{eq_ex_green_2} (junto com as condições de contorno).
